\documentclass[twocolumn,showpacs,preprintnumbers,amsmath,amssymb]{revtex4}
\usepackage{graphicx}% Include figure files
\usepackage{dcolumn}% Align table columns on decimal point
\usepackage{bm}% bold math

\begin{document}

\preprint{APS/123-QED}

\title{Exact static solutions for discrete $\phi^4$
models free of the Peierls-Nabarro barrier: \\
Discretized first integral approach}

\author{S. V. Dmitriev,$^1$ P. G. Kevrekidis,$^2$ N. Yoshikawa,$^1$ and D. J. Frantzeskakis$^3$}
\affiliation{
$^1$ Institute of Industrial Science, the University of Tokyo, Komaba, Meguro-ku, Tokyo 153-8505, Japan \\
$^2$ Department of Mathematics and Statistics, University of Massachusetts, Amherst, MA 01003-4515, USA \\
$^3$ Department of Physics, University of Athens, Panepistimiopolis, Zografos, Athens 15784, Greece
}
\date{\today}

\begin{abstract}
We propose a generalization of the discrete Klein-Gordon models
free of the Peierls-Nabarro barrier derived in Nonlinearity {\bf
12}, 1373 (1999) and Phys. Rev. E {\bf 72}, 035602(R) (2005), such
that they support not only kinks but a one-parameter set of exact
static solutions. These solutions can be obtained iteratively from
a two-point nonlinear map whose role is played by the discretized
first integral of the static Klein-Gordon field, as suggested in
J. Phys. A {\bf 38}, 7617 (2005). We then discuss some discrete
$\phi^4$ models free of the Peierls-Nabarro barrier and identify
for them the full space of available static solutions, including
those derived recently in Phys. Rev. E {\bf 72} 036605 (2005) but
not limited to them. These findings are also relevant to standing
wave solutions of
%corresponding
discrete nonlinear
Schr{\"o}dinger models. We also study stability of the obtained
solutions. As an interesting aside, we derive the list of
solutions to the continuum $\phi^4$ equation that fill the entire
two-dimensional space of parameters obtained as the continuum
limit of the corresponding space of the discrete models.
\end{abstract}


\maketitle

\section{Introduction and Setup}

Discrete nonlinear models play a very important role in many
physical applications \cite{Kivshar,Scott}. An important class of
these models consists of a few completely integrable lattice
equations, such as the Toda lattice \cite{Toda}, the
Ablowitz-Ladik lattice \cite{AL}, and the integrable sine-Gordon
lattice \cite{Orfanidis}. The fact that these lattices possess
exact soliton solutions demonstrates that, in principle,
discreteness of the host media does not preclude the propagation
of localized coherent structures. Moreover, the mobility of
soliton-like excitations in discrete media is a key issue in many
physical contexts; for instance, the mobility of dislocations, a
kind of topological solitons, is of paramount importance in the
physics of plastic deformation of metals and other crystalline
bodies \cite{Dislocations}.

A prototypical class of discrete models, relevant to a variety of
applications, consists of the Klein-Gordon dynamical lattices
\cite{Kivshar}. One of the main representatives of this family of
models is the so-called $\phi^4$ model \cite{belova}, which
features a cubic nonlinearity. This simple power law nonlinearity
renders this model a ripe testbed for studying the existence and
stability of nonlinear solutions, and comparing their properties
in continua and lattices.

In the $(1+1)$-dimensional continuum framework (and in the absence
of spatially dependent external potentials), a solution can be
shifted arbitrarily along $x$ by any $x_0$ ($x$ is the spatial
coordinate and $x_{0}={\rm const.}$), due to the existing
translational invariance. On the other hand, in the discrete
system, translational invariance is generically lost and
equilibrium static solutions exist only for a discrete rather than for a
continuum set of $x_0$ \cite{Kivshar}. Some of these equilibrium
solutions correspond to energy maxima and are unstable, while
others, corresponding to energy minima, are stable. The difference
between such maxima and minima of the energy is typically referred
to as
%the Peierls-Nabarro potential and it will be called here as
the Peierls-Nabarro barrier (PNb).
%to include the case of the
%non-Hamiltonian models with non-potential forces. (For a possible
%definition of the PNb in non-Hamiltonian lattices see discussion
%below.)

It is of particular interest to develop discretizations that do
not feature such barriers. In such cases, one might
expect that the ensuing models would be more faithful
representations of their continuum counterparts, regarding both
symmetry properties and travelling solution features (even though
there are some notes of caution that should be made; see e.g. the
discussion of \cite{aigner}).

In that vein, recently, a number of non-integrable discrete
Klein-Gordon equations free of the Peierls-Nabarro barrier
(PNb-free) has been systematically constructed. The first set of
such models which were, by construction, Hamiltonian ones, was
obtained by Speight and co-workers \cite{SpeightKleinGordon} using
the Bogomol'nyi argument \cite{Bogom}. A second successful attempt
led to the construction of momentum-conserving discretizations
\cite{PhysicaD}. It was then demonstrated, surprisingly, that the
PNb-free models of that kind conserving both energy and classical
momentum do not exist \cite{Submitted}. New PNb-free $\phi^4$
lattices were derived by Barashenkov {\it et al.}
\cite{Barashenkov}.
However, we note that for the lattices derived in
\cite{SpeightKleinGordon,Barashenkov},
%reproduce only the kink
%solution,
only the kink-type solution has been considered,
while it is well-known that
the continuum Klein-Gordon equation can support a
number of other solutions \cite{Aubry}.

On the other hand, a general approach to the construction of the
PNb-free lattices was recently reported in \cite{JPhysA}. This
approach is based on the use of the {\it discretized first
integral} (DFI) of the corresponding static field equation, and
the integration constant that enters DFI generates the
one-parameter solution space. The DFI is in fact a two-point
nonlinear algebraic equation from which the exact static solutions
of the three-point PNb-free discrete models can be found.

In this work we systematically use the DFI approach and derive the
Hamiltonian and non-Hamiltonian PNb-free Klein-Gordon lattices
supporting the one-parameter space of solutions, generalizing the
lattices offered in \cite{SpeightKleinGordon,Barashenkov}. We then
focus on the $\phi^4$ field and discuss in detail the
energy-conserving PNb-free model proposed very recently in
\cite{Saxena}, as well as the momentum-conserving model of
\cite{PhysicaD} (the two are identical in the static case). For
these models we describe the full solution space of the underlying
static problem that also includes the solutions derived earlier in
\cite{Saxena}.

Our setting is the following: We consider the Hamiltonian of the
Klein-Gordon field, $H=E_K+E_P$, with the kinetic and potential
energy functionals respectively defined as,
\begin{equation} \label{KinEn}
   E_K = \frac{1}{2} \int_{-\infty}^{\infty} \phi_{t}^2dx\,,
\end{equation}
\begin{equation} \label{PotEn}
   E_P = \frac{1}{2} \int_{-\infty}^{\infty} \left[ \phi_{x}^2
   +2V(\phi) \right]dx\,,
\end{equation}
where $\phi(x,t)$ is the unknown field and $V(\phi)$ is a given
potential function. The corresponding equation of motion is
\begin{equation} \label{KG}
   \phi_{tt} = \phi_{xx} - V'(\phi)  \equiv D(\phi(x;t))\,,
\end{equation}
where $V'(\phi)=dV/d\phi$.

Equation (\ref{KG}) will be discretized on the lattice $x=nh$,
where $n=0,\pm 1, \pm 2 ...$, and $h$ is the lattice spacing.

We would like to construct a nearest-neighbor discrete analog to
Eq. (\ref{KG}) of the form
\begin{equation} \label{KleinGordonDiscrete1}
   \ddot{\phi}_{n} = D(h,\phi_{n-1},\phi_{n},\phi_{n+1}),
\end{equation}
such that in the continuum limit $(h \rightarrow 0)$ we have
\begin{equation} \label{ContLimit}
   D(h,\phi_{n-1},\phi_{n},\phi_{n+1}) \rightarrow D(\phi),
\end{equation}
{\it and} that the solution to the three-point static problem
corresponding to Eq. (\ref{KleinGordonDiscrete1}),
\begin{equation} \label{ThreePointStatic}
   D(h,\phi_{n-1},\phi_{n},\phi_{n+1}) = 0,
\end{equation}
can be found from a reduced two-point problem
\begin{equation} \label{TwoPointStatic}
   U(h,\phi_{n-1},\phi_{n}) = 0.
\end{equation}
If this reduction is achieved, then the exact static solutions can
be constructed upon solving the algebraic Eq.
(\ref{TwoPointStatic}) iteratively, starting from arbitrary
admissible value of $\phi_{n-1}$ or $\phi_n$. Arbitrariness in the
choice of the initial condition implies that the static solution
can be placed anywhere with respect to the lattice and, for that
reason, such lattices are called translationally invariant.

Discretization Eq. (\ref{KleinGordonDiscrete1}) may result in a
non-Hamiltonian model due to the non-potential nature of the
background forces. In the latter setting, the absence of an energy
functional renders ambiguous the definition of the PN barrier,
hence we clarify this point in what follows. Suppose that we have
two equilibrium solutions, $\phi_n^{(1)}$ and $\phi_n^{(2)}$
(often, in Hamiltonian models with PNb, the second one corresponds
to a linearly unstable energy maximum while the first one to a
linearly stable energy minimum). The work done by the
inter-particle and background forces to move the $n$th particle
from the position $\phi_n^{(1)}$ to the position $\phi_n^{(2)}$ is
$W_n= \int_{\phi_n^{(1)}} ^{\phi_n^{(2)}}
D(h,\phi_{n-1},\phi_{n},\phi_{n+1}) {\rm d}\phi_n$ and the total
work performed to ``transform'' the solution $\phi_n^{(1)}$ to
$\phi_n^{(2)}$ is $W=\sum_{-\infty}^{\infty}W_n$. The PN barrier
is defined to be equal to $W$. For Hamiltonian models this
definition is equivalent to the classical one because $W$ is equal
to the energy difference between the second (possibly higher
energy unstable) configuration and the first (possibly lower
energy stable) configuration. For non-Hamiltonian models, $W_n$
will depend on the path connecting the initial and final
configurations of particles. To calculate $W$ one, therefore, has
to specify such a path. While the height of the PN barrier in a
non-Hamiltonian lattice depends on the path, the PNb-free,
non-Hamiltonian lattice can be unambiguously defined as the
translationally invariant lattice where the quasi-static
transformation between the two configurations of interest can be
done continuously through the set of {\em equilibrium}
configurations. Along the path through the equilibrium
configurations, forces acting on particles are zero and thus,
$W_n=0$ for any $n$ which, in turn, results in $W=0$. This notion
of the PNb-free lattice is also applicable to the Hamiltonian
lattices and will serve as our definition of the PNb-free models;
notice, however, that we do not overlook the mathematical
subtleties involved in this definition, including the question of
whether $W$ is zero along all realizable paths connecting
$\phi_n^{(1)}$ and $\phi_n^{(2)}$ in the non-Hamiltonian models,
among equilibrium configurations, as is the case for their
Hamiltonian siblings. These and related questions, including an
appropriate modified definition of the relevant quantities for
models such as those of the discrete nonlinear Schr{\"o}dinger
type, will be left for a future publication.

Our scope in what follows is to generalize the approach developed in
\cite{JPhysA} to  show how to construct {\it all possible}
exact static solutions of the models of \cite{JPhysA} and
\cite{Saxena} (see also \cite{PhysicaD,Barashenkov}), including
those derived in \cite{Saxena}. This will also lead us to
introduce a number of solutions (both localized and extended ones)
that, to the best of our knowledge, have not been
discussed/analyzed previously, such as the ones that will be
termed ``inverted'' (see below). We will also discuss the
stability of certain solutions among the obtained ones, for each
of the models of interest (since their stability properties are
different).

We also note in passing that while our presentation will be geared
towards the $\phi^4$ models, our results regarding the existence
of solutions can equally well be
applied to discrete equations of the nonlinear Schr{\"o}dinger
(NLS) type \cite{ablo}, such as e.g. the Ablowitz-Ladik model. In
particular, let us consider equations of the form:
\begin{eqnarray}
i \dot{\psi}_n=\frac{1}{h^2}(\psi _{n-1}-2\psi_n+\psi_{n+1}) +
f(\psi_n,\bar{\psi}_n ),
\end{eqnarray}
(where the overbar denotes complex conjugate) with $f(\psi_n
\exp(i \theta),\bar{\psi}_n \exp(-i \theta))= f(\psi_n,
\bar{\psi}_n ) \exp(i \theta)$ and $\lim_{h \rightarrow 0}
f(\psi_n, \bar{\psi}_n )= -\lambda |\psi|^2 \psi$. Then, looking
for standing wave solutions of the form $\psi_n=\exp(i \lambda t)
\phi_n$, with $\phi_n$ real, one would obtain the Klein-Gordon
static problem for the standing wave spatial profile $\phi_n$.
Hence all the discussion given below for the existence of such
solutions can be appropriately translated in the existence of
standing waves of the corresponding discrete NLS models. The
reader should be cautioned however that the stability properties
in the latter context may differ [a relevant example will be
discussed later in the text].

Our presentation will be structured as follows. In section
\ref{sec:PNpFreeKG}, we develop the DFI approach and derive
PNb-free Klein-Gordon models supporting one-parameter space of
static solutions. In section \ref{sec:PHI4}, we present some
PNb-free $\phi^4$ lattices and describe some of their basic
properties. In section \ref{ExactSolutionsDiscrete}, we discuss
the details of the construction of the general exact static
solutions of the models both in the localized (hyperbolic
function) and in the extended (general elliptic function) form,
and for both signs of the nonlinearity prefactor $\lambda$.
$\lambda=1$ corresponds to the so-called defocusing case, while
$\lambda=-1$ corresponds to the focusing case in the standard
terminology of NLS equations. Examples of the solutions are then
given in section \ref{Analysis}. In section \ref{LinearStability},
we analyze the stability of the obtained solutions. In section
\ref{Dynamics}, slow kink dynamics in the PNb-free models is
compared numerically with that in the classical $\phi^4$ model. In
section \ref{ContinuumSolutions} we give a complete list of
bounded and unbounded exact solutions to the continuum $\phi^4$
field. In section \ref{Conclusions}, we summarize our findings and
present our conclusions.

%--------------------------------------------------------------------------
\section{PNb-free Klein-Gordon lattices} \label{sec:PNpFreeKG}
%--------------------------------------------------------------------------

\subsection{Discretized first integral} \label{sec:DFI}

Following the lines of the DFI approach of Ref. \cite{JPhysA}, we
start from the first integral of the static Eq. (\ref{KG}),
\begin{eqnarray} \label{FirstIntU}
   U(x) \equiv \phi_x^2 - 2V(\phi) +C = 0 \,,
\end{eqnarray}
where $C$ is the integration constant. The first integral can also
be taken in modified forms \cite{JPhysA}, e.g., as
\begin{eqnarray} \label{tildeFirstIntv}
   \tilde{v}(x) \equiv p\left[g(\phi_x^2) - g(2V-C)\right] = 0 \,,
\end{eqnarray}
or as
\begin{eqnarray} \label{tildeFirstIntw}
   \tilde{w}(x) \equiv p\left[g(\phi_x^2 +C) - g(2V)\right] = 0 \,,
\end{eqnarray}
where $p$ and $g$ are continuous functions and $p(0)=0$.
Note that $\tilde{v}(x)$ and $\tilde{w}(x)$ are equivalent only if
$g(\xi)=\xi$. We will consider the case of $p(\zeta)=\zeta$,
$g(\xi)=\xi$, i.e., the unchanged form of the first integral, Eq.
(\ref{FirstIntU}), together with the case of $p(\zeta)=\zeta$,
$g(\xi)=\sqrt{\xi}$, for which we have the two possibilities
\begin{eqnarray} \label{FirstIntv}
   v(x) \equiv \pm \phi_x - \sqrt{2V(\phi)-C} = 0 \,,
\end{eqnarray}
and
\begin{eqnarray} \label{FirstIntw}
   w(x) \equiv \pm \sqrt{\phi_x^2+C} - \sqrt{2V(\phi)} = 0 \,,
\end{eqnarray}
of which only the first one will be discussed below.

We then construct the DFIs corresponding to Eq. (\ref{FirstIntU})
and Eq. (\ref{FirstIntv}), which are respectively given by
\begin{eqnarray} \label{TwoPointMapU}
   U(h,\phi_{n-1},\phi_{n}) \equiv \frac{(\phi_{n}
   -\phi_{n-1})^2}{h^2} \nonumber \\
   -2V(\phi_{n-1},\phi_{n}) + C = 0,
\end{eqnarray}
\begin{eqnarray} \label{TwoPointMapv}
   v(h,\phi_{n-1},\phi_{n}) \equiv \pm \frac{\phi_{n}
   -\phi_{n-1}}{h}  \nonumber \\
   -\sqrt{2V(\phi_{n-1},\phi_{n}) - C} = 0,
\end{eqnarray}
where we demand that $V(\phi_{n-1},\phi_{n}) \rightarrow V(\phi)$
in the continuum limit ($h \rightarrow 0$).

\subsection{Momentum-conserving PNb-free lattice} \label{sec:MomCons}

First we construct a PNb-free Klein-Gordon lattice using the
unchanged form of the first integral, Eq. (\ref{FirstIntU}), and
corresponding DFI, Eq. (\ref{TwoPointMapU}). Calculating $dU/dx$
and multiplying the result by $(dx/d\phi)/2$, we find
\begin{eqnarray} \label{D1}
   \frac{1}{2}\frac{dU}{d\phi} = D(x).
\end{eqnarray}
Discretizing the left-hand side of Eq. (\ref{D1}) we obtain the
lattice Klein-Gordon equation
\begin{eqnarray} \label{D1discr}
   \ddot{\phi}_n = \frac{U(h,\phi_{n},\phi_{n+1})
   - U(h,\phi_{n-1},\phi_{n})}{\phi_{n+1}-\phi_{n-1}},
\end{eqnarray}
whose static solutions can be found from the two-point problem,
Eq. (\ref{TwoPointMapU}), and thus, the lattice is PNb-free.

The lattice of this type was first derived in \cite{PhysicaD} where it
was also demonstrated that it conserves the momentum
\begin{eqnarray}
P=\sum_n \dot{\phi}_n (\phi_{n+1}-\phi_{n-1}). \label{mom}
\end{eqnarray}

The integration constant $C$ that
%enters
appears in Eq. (\ref{TwoPointMapU})
cancels out in Eq. (\ref{D1discr}). This means that all kinds of
static solutions derived from Eq. (\ref{TwoPointMapU}) for
different $C$ (kink solution corresponds to $C=0$) will be the
static solutions to one and the same Klein-Gordon lattice, Eq.
(\ref{D1discr}), since it does not depend on $C$.

The right-hand side of Eq. (\ref{D1discr}) becomes non-singular
when the discretization of the potential $V(\phi)$ is a polynomial
function having symmetry $V(\phi_{n-1}, \phi_{n}) = V(\phi_{n},
\phi_{n-1})$. The most general expression of this type was given
in \cite{Submitted}.

\subsection{Energy-conserving PNb-free lattice} \label{sec:Hamilt}

To construct Hamiltonian PNb-free lattices we discretize not the
equation of motion, Eq. (\ref{KG}), but the Hamiltonian, Eq.
(\ref{KinEn}) and Eq. (\ref{PotEn}). We now use the modified first
integral in the form of Eq. (\ref{FirstIntv}) with the upper sign
and rewrite the potential energy functional, Eq. (\ref{PotEn}), as
follows,
\begin{equation} \label{PotEnFI}
   E_P =\frac{1}{2} \int_{-\infty}^{\infty} \left( \left[v(x)\right]^2
   +2\phi_x\sqrt{2V(\phi)-C} \right)dx\,,
\end{equation}
where we omitted the constant term. Discretizing the kinetic
energy, Eq. (\ref{KinEn}), and the potential energy, Eq.
(\ref{PotEnFI}), we obtain the discrete Hamiltonian
\begin{eqnarray} \label{HamiltonianDiscr}
   {\cal H} &=&\frac{1}{2} \sum_{n} \Big\{ \dot{\phi}_n^2
   +\left[v(h,\phi_{n-1},
   \phi_n)\right]^2 \nonumber \\
   &+& 2\frac{\phi_n-\phi_{n-1}}{h}\sqrt{2V(\phi_{n-1},\phi_n)-C}\Big\}\,.
\end{eqnarray}
If the background potential is discretized as suggested in
\cite{SpeightKleinGordon},
\begin{eqnarray} \label{PNpFreeCondition}
   \sqrt {2V\left( {\phi_{n-1},\phi_n} \right) -C}  &=&
   \frac{{G(\phi_n) - G(\phi_{n-1})}}{{\phi_n - \phi_{n-1}}}, \nonumber \\
   {\rm where} \quad G'\left( \phi  \right) &=& \sqrt {2V(\phi)-C},
\end{eqnarray}
then the last term of the Hamiltonian Eq. (\ref{HamiltonianDiscr})
reduces to $(2/h)[G(\phi_n) - G(\phi_{n-1})]$ and it disappears in
the telescopic summation. With the choice of Eq.
(\ref{PNpFreeCondition}), the equations of motion derived from Eq.
(\ref{HamiltonianDiscr}) are
\begin{eqnarray} \label{HamiltPNpFree}
  \ddot{\phi}_n=-\hat{v}(h,\phi_{n-1},\phi_n)\frac{\partial }{{\partial \phi_n}}
  \hat{v}(h,\phi_{n-1},\phi_n) \nonumber \\
  -\hat{v}(h,\phi_{n},\phi_{n+1})\frac{\partial }{{\partial \phi_n}}
  \hat{v}(h,\phi_{n},\phi_{n+1}),
\end{eqnarray}
where, according to Eq. (\ref{TwoPointMapv}) and Eq.
(\ref{PNpFreeCondition}),
\begin{eqnarray} \label{TwoPointMapv1}
   \hat{v}(h,\phi_{n-1},\phi_{n}) = \frac{\phi_{n}
   -\phi_{n-1}}{h}  %\nonumber \\
   -\frac{{G(\phi_n) - G(\phi_{n-1})}}{{\phi_n - \phi_{n-1}}}.
\end{eqnarray}
This lattice conserves the Hamiltonian (total energy)
\begin{eqnarray} \label{HamiltonianDiscr1}
   \hat{{\cal H}} = \frac{1}{2}\sum_{n} \Big\{ \dot{\phi}_n^2
   +\left[\hat{v}(h,\phi_{n-1},
   \phi_n)\right]^2\Big\}\,.
\end{eqnarray}
Obviously, static solutions to the lattice Eq.
(\ref{HamiltPNpFree}) can be found from the two-point DFI
$\hat{v}(h,\phi_{n-1},\phi_n)=0$ and it is also clear that the
Hamiltonian Eq. (\ref{HamiltonianDiscr1}) is PNb-free.

The energy-conserving lattice involves the integration constant
$C$ (through the function $G$) and this is different from what we
had for the momentum-conserving model. The special case of $C=0$ yields
the original model by Speight \cite{SpeightKleinGordon}, which
supports a kink solution. Models with $C \neq 0$ describe
solutions different from the kink solution.

\subsection{ Possible generalizations } \label{sec:Generalizations}

Suppose that we use the same function $V(\phi_{n-1},\phi_{n})$ in
the DFIs Eq. (\ref{TwoPointMapU}) and Eq. (\ref{TwoPointMapv}) to
construct different $D_i(h,\phi_{n-1},\phi_{n},\phi_{n+1})$ terms.
Then equations $U(h,\phi_{n-1},\phi_{n})=0$ and
$v(h,\phi_{n-1},\phi_{n})=0$ are equivalent. Linear combination of
those terms can be used to write the following PNb-free
Klein-Gordon model
\begin{equation} \label{KleinGordonDiscrAll}
   \ddot{\phi_{n}} = \sum_i b_iD_i(h,\phi_{n-1},\phi_{n},\phi_{n+1}),
\end{equation}
where the constant coefficients satisfy the continuity constraint
\begin{equation} \label{ContConstr}
   \sum_ib_i\rightarrow 1 \quad {\rm when}\quad h \rightarrow 0,
\end{equation}
assuming that $b_i$ can depend on $h$. Static solutions of Eq.
(\ref{KleinGordonDiscrAll}) can be found iteratively from Eq.
(\ref{TwoPointMapU}) or Eq. (\ref{TwoPointMapv}).

The model of Eq. (\ref{KleinGordonDiscrAll}) can be generalized in
a number of ways. For example, one can append terms which
disappear in the continuum limit and  ones that vanish upon
substituting $U(h,\phi_{n-1},\phi_{n})=0$. Let us call such terms
as $O$-terms. Furthermore, any term $D_i(h,\phi_{n-1}, \phi_{n},
\phi_{n+1})$ can be modified by multiplying by a continuous
function $e(h,\phi_{n-1},\phi_{n},\phi_{n+1})$, which never
vanishes and whose continuum limit is unity. Such multiplication
will not change either the continuum limit, or the static
solutions of the model.

Particularly we will study the PNb-free model of the form
\begin{eqnarray} \label{BOP}
   \ddot{\phi}_n = e(h,\phi_n)\frac{U(h,\phi_{n},\phi_{n+1})
   - U(h,\phi_{n-1},\phi_{n})}{\phi_{n+1}-\phi_{n-1}} \nonumber \\
   +\rho\Big[(\phi_{n}-\phi_{n-1})v(\phi_{n},\phi_{n+1}) \nonumber \\
   -(\phi_{n+1}-\phi_{n})v(\phi_{n-1},\phi_{n})\Big],
\end{eqnarray}
which is the lattice Eq. (\ref{D1discr}) modified by a multiplier
$e(h,\phi_n)$ and augmented with the $O$-term with arbitrary
coefficient $\rho$. This $O$-term was constructed to fit to the
$I_3$ invariant offered in \cite{Barashenkov} for the discretization
of the $\phi^4$ field.

Finally we note that expressions similar to Eq. (\ref{D1}) can be
derived from derivatives $d\tilde{v}(x)/dx$ and $d\tilde{w}(x)/dx$
and they can produce new PNb-free models.

%--------------------------------------------------------------------------
\section{PNb-free discrete $\phi^4$ models} \label{sec:PHI4}
%--------------------------------------------------------------------------

To give examples of Klein-Gordon lattices, we will discretize the
well-known $\phi^4$ field with the potential
\begin{equation}
V(\phi)=\frac{\lambda}{4}(1-\phi^2)^2, \label{Potential}
\end{equation}
where the parameter $\lambda = \pm 1$. The corresponding equation
of motion is
\begin{equation}
\phi _{tt}  = \phi _{xx} +\lambda\left(\phi - \phi^3\right).
\label{KleinGordon}
\end{equation}
The above, so-called, $\phi^4$ equation supports moving periodic
solutions that can be expressed in terms of the Jacobi elliptic
functions. The latter can be reduced to localized hyperbolic
function solutions (when the elliptic modulus $m=1$). Bounded
solutions of this sort were discussed in the context of structural
phase transitions \cite{Aubry} and were used as the starting point
for derivation of exact solutions to a discrete $\phi^4$ model
\cite{Saxena}. In section \ref{ContinuumSolutions} we complete the
list of the solutions presented in \cite{Saxena,Aubry} by
providing also the set of unbounded solutions of the present
model.

The simplest discretization of Eq. (\ref{KleinGordon}) is of the
form
\begin{equation}
{\ddot \phi}_n  = \frac{1}{h^2}(\phi _{n-1}-2\phi_n+\phi_{n+1}) +
\lambda \left(\phi_n - \phi_n^3\right),
\label{KleinGordonDiscrete}
\end{equation}
and it possesses a PN barrier.
However, as mentioned above, one can construct PNb-free discrete
Klein-Gordon models by discretizing the nonlinear term
$V^{\prime}(\phi)$ typically over three neighboring points,
$V^{\prime}(\phi) \rightarrow V^{\prime} (\phi_{n-1}, \phi_{n},
\phi_{n+1})$, in contrast to the classical discretization
$V^{\prime} (\phi) \rightarrow V^{\prime} (\phi_n)$. The
three-point discretization may be physically meaningful in some
settings \cite{Coulomb}, but is also interesting from the more
fundamental point of view of developing PNb-free discretizations
and obtaining analytically (or semi-analytically) explicit
waveforms of their solutions.

We discretize the potential of Eq. (\ref{Potential}) as follows
\begin{eqnarray} \label{Phi4potentialDiscr}
   V(\phi_{n-1}, \phi_{n})= \frac{\lambda}{4}
   \left( 1-\phi_{n-1}\phi_{n} \right)^2.
\end{eqnarray}
Then, the DFIs, Eqs. (\ref{TwoPointMapU}) and (\ref{TwoPointMapv}),
become respectively,
\begin{eqnarray} \label{FirstIntDiscrete}
   U(h,\phi_{n-1},\phi_{n}) &\equiv& \frac{(\phi_{n}
   -\phi_{n-1})^2}{h^2} \nonumber \\
   &-&\frac{\lambda}{2}
   \left( 1-\phi_{n-1}\phi_{n} \right)^2 + C = 0,
\end{eqnarray}
\begin{eqnarray} \label{TwoPointMapvKG}
   v(h,\phi_{n-1},\phi_{n}) &\equiv & \frac{\phi_{n}
   -\phi_{n-1}}{h}  \nonumber \\
   &-&\sqrt{\frac{\lambda}{2}
   \left( 1-\phi_{n-1}\phi_{n} \right)^2 - C} = 0.
\end{eqnarray}
It has been demonstrated that at $C=0$ Eq.
(\ref{FirstIntDiscrete}) and Eq. (\ref{TwoPointMapvKG}) support
kink solutions \cite{SpeightKleinGordon,Barashenkov}. For $C\neq
0$ they support solutions different from kink.

Let us denote
\begin{eqnarray}
\Lambda=\lambda h^2,\,\,\,\,\,\,\,\,\,\,\,\,\,\,\,\,\,
\tilde{C}=Ch^2. \label{Notation}
\end{eqnarray}

Substituting Eq. (\ref{FirstIntDiscrete}) and Eq.
(\ref{TwoPointMapvKG}) into Eq. (\ref{BOP}) with $e(h,\phi_n)=1$
we
%come to
arrive at the following PNb-free $\phi^4$ lattice
\begin{eqnarray} \label{Barashenkov1}
   \ddot{\phi}_n &=& \Delta_2\phi_n + \lambda\phi_n -\frac{\lambda}{2}\phi_n^2
   (\phi_{n-1}+\phi_{n+1}) \nonumber \\
   &+&\rho(\phi_{n+1}-\phi_n)
   \sqrt{\frac{\lambda}{2}(1-\phi_{n-1}\phi_n)^2-C} \nonumber \\
   &-&\rho(\phi_{n}-\phi_{n-1})
   \sqrt{\frac{\lambda}{2}(1-\phi_{n}\phi_{n+1})^2-C}.
\end{eqnarray}
Equation (\ref{Barashenkov1}) gives a one-parameter $(C)$ family
of PNb-free models. For the special case of $C=0$ and $\lambda=1$
we get one of the lattices derived in \cite{Barashenkov},
$\ddot{\phi}_n = (1+h^2\rho/\sqrt{2})\Delta_2\phi_n + \phi_n
+(\rho/\sqrt{2}-1/2)\phi_n^2 (\phi_{n-1}+\phi_{n+1}) -\rho\sqrt{2}
\phi_{n-1}\phi_{n}\phi_{n+1}$, and this lattice supports the kink
solution. Lattices with $C\neq 0$ support solutions different from
the kink.

For $\rho=0$ we get from Eq. (\ref{Barashenkov1}) the
$C$-independent model
\begin{eqnarray}
{\ddot \phi_n}&=&\frac{1}{h^2}(\phi _{n-1}-2\phi_n+\phi_{n+1})  \nonumber \\
&+& \lambda \phi_n -\frac{\lambda}{2} \phi_n^2 \left( \phi_{n-1} +
\phi_{n+1} \right).
\label{PhysicaD}
\end{eqnarray}
This non-Hamiltonian PNb-free $\phi^4$ model conserves the
momentum Eq. (\ref{mom}) and it will be referred to as the
momentum-conserving (MC) model.

Substituting Eq. (\ref{FirstIntDiscrete}) into Eq. (\ref{BOP})
with $e(h,\phi_n)=1/(1- \Lambda \phi_n^2/2)$ and $\rho=0$ we
obtain another $C$-independent PNb-free model discovered in
\cite{Saxena},
\begin{equation}
\ddot{\phi}_n  = \frac{1}{h^2}(\phi _{n-1}-2\phi_n+\phi_{n+1}) +
\frac{\lambda \left(\phi_n - \phi_n^3\right)}{1 - \Lambda
\phi_n^2/2}. \label{Saxena}
\end{equation}
This model will be called the energy-conserving (EC) model because
it possesses the Hamiltonian \cite{Saxena}
\begin{eqnarray}
{\cal H} =\frac{1}{2}\sum_{n} \Big[ {\dot \phi}_n^2 +
\frac{(\phi_n-\phi_{n-1})^2}{h^2}+ V(\phi_n)\Big], \label{calh}
\end{eqnarray}
where the potential $V(\phi_n)$ is given by
\begin{eqnarray}
V(\phi_n)=-\frac{1}{h^2} \left(\phi_n^2 +
\frac{2-\Lambda}{\Lambda}\ln \left|1 - \frac {\Lambda
\phi_n^2}{2}\right| \right). \label{HamSaxena}
\end{eqnarray}
\begin{figure}
\includegraphics{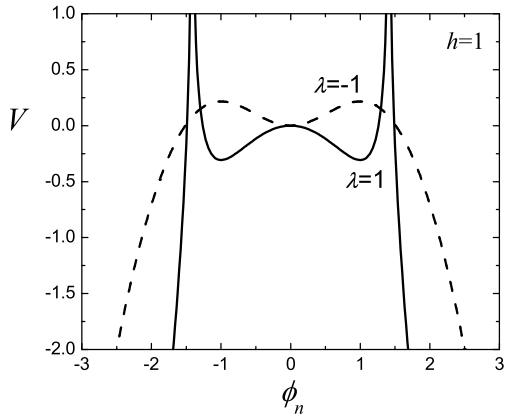}
\caption{The on-site potential of the EC model of Eq. (\ref{Saxena}),
$V(\phi_n)$, defined by Eq. (\ref{HamSaxena}), for $h=1$ and
$\Lambda=\lambda=1$ (solid line) and $\Lambda =\lambda=-1$ (dashed line). For
$\lambda<0$ the potential is smooth and it has one minimum at
$\phi_n=0$ and two maxima at $\phi_n=\pm 1$. For $\lambda>0$ the
potential has two minima at $\phi_n=\pm 1$ and a maximum at
$\phi_n=0$; as $\phi_n\rightarrow \pm \sqrt{2/\Lambda}$ the
potential $V(\phi_n) \rightarrow + \infty$.}
\label{Figure1}
\end{figure}

In Fig. \ref{Figure1} we plot the potential $V(\phi_n)$ for $h=1$ (i.e., $\Lambda=\lambda$) in both cases
$\lambda=1$ (solid line) and $\lambda=-1$ (dashed line). For
$\lambda<0$ the potential is smooth and it has one minimum at
$\phi_n=0$ and two maxima at $\phi_n=\pm 1$. For $\lambda>0$ the
potential has two minima at $\phi_n=\pm 1$ and a maximum at
$\phi_n=0$; note that in the limit $\phi_n\rightarrow \pm \sqrt{2/\Lambda}$, the
potential $V(\phi_n) \rightarrow + \infty$.

It is not possible to plot an analog of Fig. \ref{Figure1} for the
MC model since the relevant background forces are of many-body
type (i.e., involve nearest neighbors) and are non-potential.

PNb-free models given by Eq. (\ref{Barashenkov1}), Eq.
(\ref{PhysicaD}), and Eq. (\ref{Saxena}) have exactly same static
solutions derivable from DFI Eq. (\ref{FirstIntDiscrete}).

\section{Exact static solutions for discrete models}
\label{ExactSolutionsDiscrete}

\subsection{Solutions from nonlinear map}

To find {\it all} static solutions to the PNb-free models of Eq.
(\ref{Barashenkov1}), Eq. (\ref{PhysicaD}), and Eq. (\ref{Saxena})
we solve the DFI of Eq. (\ref{FirstIntDiscrete}):
\begin{eqnarray}
\phi_n=\frac{(2-\Lambda) \phi_{n-1} \pm\sqrt {\cal
D}}{2-\Lambda
\phi_{n-1}^2}, \nonumber \\
{\cal D}=2\Lambda\left( 1-\phi_{n-1}^2 \right)^2 + 2
\tilde{C}\left(\Lambda\phi_{n-1}^2 -2 \right), \label{Pulse}
\end{eqnarray}
where $\phi_n$ and $\phi_{n-1}$ can be interchanged due to the
symmetry of the equation. Starting from any admissible ``initial''
value $\phi_0$, by iterating Eq. (\ref{Pulse}) and its counterpart
written as $\phi_{n-1}=f(\phi_n)$, one can construct recurrently
the static solution to both the MC model of Eq. (\ref{PhysicaD})
and the EC model of Eq. (\ref{Saxena}), or to a linear combination
thereof. Arbitrariness in the choice of $\phi_0$ implies the
absence of PNb in these models, which has been also demonstrated
in \cite{Saxena}.

As can be seen from Eq. (\ref{Pulse}), once the values of
$\tilde{C}$ and $\Lambda$ are fixed, there are certain
restrictions on the choice of the values of $\phi_0$. In
particular, inadmissible initial values are those for which the
denominator becomes zero, i.e., $\phi_0 \neq \pm \sqrt{2/\Lambda}$
for $\lambda>0$. An exceptional case is that of $\Lambda=2$,
$\tilde{C}=0$ when an arbitrary sequence of $\pm 1$ is a solution
of Eq. (\ref{FirstIntDiscrete}). Inadmissible values of $\phi_0$
are also ones for which ${\cal D}<0$. The condition ${\cal D}=0$
leads to a biquadratic algebraic equation determining the borders
of admissible region; the roots of this equation are:
\begin{eqnarray}
(\phi_{0}^2)_{1,2}=1-\frac{\tilde{C}}{2} \pm\sqrt{ \frac{\tilde{C}}{4}
\left( \tilde{C} -4 + \frac{8} {\Lambda} \right)}.
\label{Borders}
\end{eqnarray}
Let us introduce the following notations for these roots:
\begin{eqnarray}
F_1=-F_3=\sqrt{(\phi_{0}^2)_{1}}\,,\,\,\,\,\,
F_2=-F_4=\sqrt{(\phi_{0}^2)_{2}}\,. \label{RootNotation}
\end{eqnarray}

\begin{figure}
\includegraphics{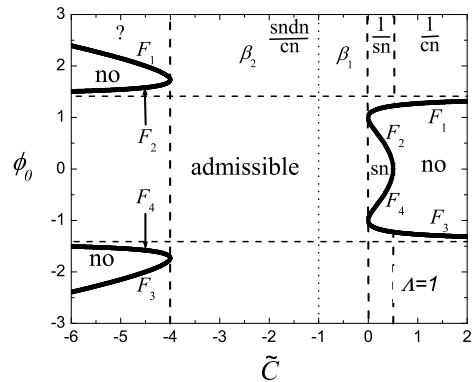}
\caption{Admissible region for the ``initial'' value $\phi_{0}$ in
the nonlinear map Eq. (\ref{Pulse}) for different values of $\tilde{C}$ at
$\Lambda=1$. There are three inadmissible regions marked with
``no'' and two inadmissible lines $\phi_{0} \neq \pm
\sqrt{2/\Lambda}$ (horizontal dashed lines). On these lines the
on-site potential $V(\phi_n)$ diverges (see Fig. \ref{Figure1}).
$F_i$ ($i=1,\dots,4$) designate different branches of the borders
of the admissible regions [see Eq. (\ref{Borders}) and Eq.
(\ref{RootNotation})]. Roots $F_1$ and $F_2$ merge at
$\tilde{C}=0$ and at $\tilde{C}=4-8/\Lambda$. Roots $F_2$ and
$F_4$ merge at $\tilde{C}=\Lambda/2$. Vertical dashed lines
separate regions with different Jacobi elliptic function
solutions. Vertical dotted line, situated (for the chosen
parameters) at $\tilde{C}^{\ast} \approx -1.00$, divides the
region of the ${\rm sndn/cn}$ solution into two portions
corresponding to two roots of the first equation in Eq.
(\ref{p1m1p1d}).}
\label{Figure2}
\end{figure}

\begin{figure}
\includegraphics{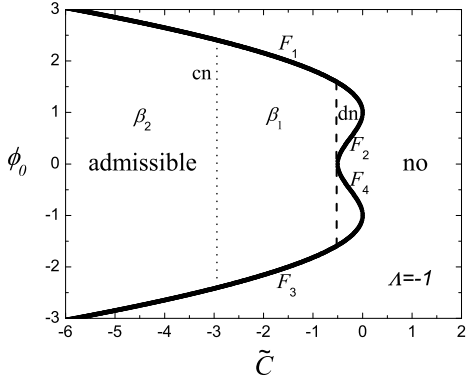}
\caption{Same as in Fig. \ref{Figure2} but for $\Lambda=-1$. Roots
$F_1$ and $F_2$ merge at $\tilde{C}=0$, and roots $F_2$ and $F_4$
merge at $\tilde{C}=\Lambda/2$. The ${\rm dn}$ solution, Eq.
(\ref{n0n0p1d}), is defined for $\Lambda/2<\tilde{C}<0$. The ${\rm
cn}$ solution, Eq. (\ref{n0p1n0d}), is defined for
$\tilde{C}<\Lambda/2$, and this region is divided into two parts,
$\beta_1$ and $\beta_2$, each corresponding to a particular root
of the first equation in Eq. (\ref{n0p1n0d}). The border between
these two regions is shown by the dotted line situated (for the
chosen parameters) at $\tilde{C}^{\ast} \approx -2.96$.}
\label{Figure3}
\end{figure}

The admissible regions for the values of $\phi_{0}$ of the
nonlinear map Eq. (\ref{Pulse}) are shown in Fig. \ref{Figure2}
for different values of $\tilde{C}$ at $\Lambda=1$. The corresponding result
for $\Lambda=-1$ is presented in Fig. \ref{Figure3}. These graphs
present a road map for constructing the various possible solutions
of the above models.

The symmetry of Eq. (\ref{FirstIntDiscrete}) suggests that the
topology of the admissible regions is such that once started from
an admissible value of $\phi_{0}$, one cannot leave the admissible
region iterating Eq. (\ref{Pulse}), so that the static solution
will surely be constructed for the whole chain. This is so because
Eq. (\ref{Pulse}) serves for calculating both back and forth
points of the map, and if one is admissible, the other one is also
admissible.

Equation (\ref{Pulse}) possesses two roots, which means that for an
admissible initial condition one can construct two different
solutions, e.g., a kink and an antikink. When iterating, to keep
moving along the same solution, one must take $\phi_{n}$ different
from $\phi_{n-2}$ [if the roots of Eq. (\ref{Pulse}) are
different]. Indeed, setting in the three-point static problem, Eq.
(\ref{PhysicaD}), $\phi_{n-1}=\phi_{n+1}$, we find
\begin{eqnarray}
\phi_{n-1}=\frac{(2-\Lambda)\phi_n}{2-\Lambda\phi_n^2}.
\label{Check}
\end{eqnarray}
Comparing this with Eq. (\ref{Pulse}), it is readily seen that the
equilibrium in the three-point equation in the case of
$\phi_{n-1}=\phi_{n+1}$ can be achieved only if ${\cal D}=0$ and
the two roots coincide. As mentioned above, the latter requirement
is equivalent to the condition that $\phi_n$ is on the border of
the admissible region.

As it can be deduced from Eq. (\ref{Borders}), the topology of the
admissible regions presented in Fig. \ref{Figure2} and Fig.
\ref{Figure3} does not significantly change among different but
positive and among different but negative values of $\Lambda$
respectively; of course, one has to exclude the particular case of
$\Lambda=2$ and also the continuum limit (see Sec.
\ref{ContinuumSolutions}). Here we will not discuss in detail the
case of extremely high discreteness $(\Lambda>2)$, even though the
analysis of this case does not present any additional
difficulties.

To summarize, inside the admissible region, $(\tilde{C},\phi_{0})$
(shown in Figs. \ref{Figure2} and \ref{Figure3}), the nonlinear
map of Eq. (\ref{Pulse}) generates static solutions to the MC
model of Eq. (\ref{PhysicaD}) and to the EC model of Eq.
(\ref{Saxena}).

\subsection{Jacobi elliptic function solutions}

Static solutions to the discrete PNb-free models of Eq.
(\ref{PhysicaD}) and Eq. (\ref{Saxena}) have been reported
\cite{Saxena} in the form of the Jacobi elliptic functions, ${\rm
sn,cn,}$ and ${\rm dn}$ \cite{SpecialFunctions}. Below, we will
derive such solutions.

The general form of the solutions is
\begin{eqnarray}
\phi_n&=&\pm A {\rm sn}^q(Z,m) {\rm cn}^r(Z,m) {\rm dn}^s(Z,m) ,
\nonumber \\
Z&=&\beta h \left(n+x_0\right), \label{EllipticDiscrete}
\end{eqnarray}
where $0 \le m \le 1$ is the modulus of the Jacobi elliptic
functions, $A$ and $\beta$ are the parameters of the solution, and
$x_0$ is the arbitrary initial position. Finally the integers $q,r,s$ specify a
particular form of the solution.

In the limit of $m=1$, Eq. (\ref{EllipticDiscrete}) reduces to the
hyperbolic function form
\begin{eqnarray}
\phi_n=\pm A {\tanh}^q(Z) {\cosh}^{-r-s}(Z) ,
\label{HyperbolicDiscrete}
\end{eqnarray}
and, in the limit of $m=0$, to the trigonometric function form
\begin{eqnarray}
\phi_n=\pm A {\sin}^{q}(Z) {\cos}^{r}(Z) .
\label{TrigonomDiscrete}
\end{eqnarray}

Substituting the ansatz Eq. (\ref{EllipticDiscrete}) into Eq.
(\ref{FirstIntDiscrete}) and equalizing the coefficients in front
of similar terms we find that it can be satisfied for a limited
number of combinations of integer powers $q,r,s$. Solving the
equations for the coefficients we relate the parameters of the
solution Eq. (\ref{EllipticDiscrete}), $\beta$ and $A$, to
$\Lambda$ and $m$  and also find the relation between $C$, that
enters Eq. (\ref{FirstIntDiscrete}), and $m$.

For some of the sets $(q,r,s)$, e.g., for $(q,r,s)=(1,-1,0)$ and
for $(q,r,s)=(1,-1,-1)$, we obtain imaginary amplitude $A$ in the
whole range of parameters.

Essentially different, real amplitude solutions described by Eq.
(\ref{EllipticDiscrete}), have the following form and are characterized by the following parameters:

The ${\rm sn}$ solution, $(q,r,s)=(1,0,0)$,
\begin{eqnarray}
{\rm cn}(\beta h){\rm dn}(\beta h) = 1- \frac{\Lambda}{2},
\,\,\,\,\,\,\,\, A=\sqrt{\frac{2m}{\Lambda}}{\rm sn}(\beta h),
\nonumber \\
\tilde{C}=\frac{\Lambda}{2} \left(1 - \frac{A^4}{m} \right),
\,\,\,\,\,\,\,\,\,\,\,0<\tilde{C}<\frac{\Lambda}{2}.
\label{p1n0n0d}
\end{eqnarray}

The ${\rm cn}$ solution, $(q,r,s)=(0,1,0)$,
\begin{eqnarray}
\frac{{\rm cn}(\beta h)}{{\rm dn}^2(\beta h)} = 1-
\frac{\Lambda}{2}, \,\,\,\,\,\,\,\,
A=\sqrt{\frac{-2m}{\Lambda}}\frac{{\rm sn}(\beta h)} {{\rm
dn}(\beta h)},
\nonumber \\
\tilde{C}=\Lambda \frac{(1-A^2)^2}{2-\Lambda A^2},
\,\,\,\,\,\,\,\,\,\,\,-\infty <\tilde{C}< \frac{\Lambda}{2}.
\label{n0p1n0d}
\end{eqnarray}

The ${\rm dn}$ solution, $(q,r,s)=(0,0,1)$,
\begin{eqnarray}
\frac{{\rm dn}(\beta h)}{{\rm cn}^2(\beta h)} = 1-
\frac{\Lambda}{2}, \,\,\,\,\,\,\,\,
A=\sqrt{\frac{-2}{\Lambda}}\frac{{\rm sn}(\beta h)} {{\rm
cn}(\beta h)},
\nonumber \\
\tilde{C}=\Lambda \frac{(1-A^2)^2}{2-\Lambda A^2},
\,\,\,\,\,\,\,\,\,\,\,\frac{\Lambda}{2}<\tilde{C}<0.
\label{n0n0p1d}
\end{eqnarray}

The $1/{\rm sn}$ solution, $(q,r,s)=(-1,0,0)$,
\begin{eqnarray}
{\rm cn}(\beta h){\rm dn}(\beta h) = 1- \frac{\Lambda}{2},
\,\,\,\,\,\,\,\, A=\sqrt{\frac{2}{\Lambda}}{\rm sn}(\beta h),
\nonumber \\
\tilde{C}=\frac{\Lambda}{2}(1 - m A^4),
\,\,\,\,\,\,\,\,\,\,\,0<\tilde{C}<\frac{\Lambda}{2}.
\label{m1n0n0d}
\end{eqnarray}

The $1/{\rm cn}$ solution, $(q,r,s)=(0,-1,0)$,
\begin{eqnarray}
\frac{{\rm cn}(\beta h)}{{\rm dn}^2(\beta h)} = 1-
\frac{\Lambda}{2}, \,\,\,\,\,\,\,\, A=\sqrt{ \frac{2(1-m)}
{\Lambda}}\frac{{\rm sn}(\beta h)} {{\rm dn}(\beta h)},
\nonumber \\
\tilde{C}=\frac{\Lambda}{2}\left(1+ \frac{mA^4} {1-m} \right),
\,\,\,\,\,\,\,\,\,\,\,\frac{\Lambda}{2} <\tilde{C}< \infty.
\label{n0m1n0d}
\end{eqnarray}

The ${\rm sndn/cn}$ solution, $(q,r,s)=(1,-1,1)$,
\begin{eqnarray}
\frac{m{\rm cn}^4(\beta h)+1-m}{{\rm cn}^2(\beta h)} = 1-
\frac{\Lambda}{2}, \,\, A=\sqrt{ \frac{2} {\Lambda}}\frac{{\rm
sn}(\beta h){\rm dn}(\beta h)} {{\rm cn}(\beta h)},
\nonumber \\
\tilde{C}=\frac{\Lambda}{2}\left(1- A^4 \right),
\,\,\,\,\,\,\,\,\,\,\,4- \frac{8}{\Lambda}< \tilde{C}
<0.\,\,\,\,\,\,
\label{p1m1p1d}
\end{eqnarray}

It is worth making the following remarks:

The solutions shown in Eq. (\ref{n0p1n0d}) and Eq. (\ref{n0n0p1d}) have
real amplitudes for $\Lambda <0$ while the others for $\Lambda>0$.

The solutions should be interpreted in the following form. For a
given $\Lambda$, one can find $\beta$ by solving the first
equation in Eqs. (\ref{p1n0n0d}-\ref{p1m1p1d}), and then $A$ from
the second one. Substituting these values in Eq.
(\ref{EllipticDiscrete}) results in the static solutions of the
original discrete model.

The expressions for $\tilde{C}$ in Eqs.
(\ref{p1n0n0d}-\ref{p1m1p1d}) link the elliptic Jacobi function
solutions and the solution in the form of the nonlinear map, Eq.
(\ref{Pulse}). As for the other free parameter of the solutions
Eq. (\ref{p1n0n0d}-\ref{p1m1p1d}), the arbitrary shift $x_0$, its
counterpart in the nonlinear map, Eq. (\ref{Pulse}), is
effectively the initial value $\phi_0$.

The solutions of Eqs. (\ref{p1n0n0d}-\ref{p1m1p1d}) can be expressed
in a number of other forms using the well-known identities for the
Jacobi elliptic functions \cite{SpecialFunctions}. For example,
shifting the argument by a quarter period, one can transform the
${\rm sn}$ solution to the form of ${\rm cn/dn}$, or, applying the
ascending Landen transformation, to the form of ${\rm sncn/dn}$.
Mathematically, these three expressions look as different members
of Eq. (\ref{EllipticDiscrete}), but physically they are
indistinguishable.

\section{Analysis of static solutions}
\label{Analysis}

In what follows, we analyze the static solutions derivable from
Eq. (\ref{Pulse}) discussing their relation to Eqs.
(\ref{p1n0n0d})-(\ref{p1m1p1d}). Since the topology of the
admissible regions is different for positive and negative
$\Lambda$, these two cases are studied separately.

\subsection{$\Lambda>0$ case.}

In the ${\rm sn}$ solution, Eq. (\ref{p1n0n0d}), when the elliptic modulus
increases from its smallest value $m=0$ to its largest value
$m=1$, the amplitude $A$ increases from $0$ to $1$, since $A=F_2$
[see Eq. (\ref{Borders}) and Eq. (\ref{RootNotation})]. As a
result, the parameter $\tilde{C}$ monotonically decreases from
$\Lambda/2$ to $0$. Thus, the ${\rm sn}$ solution is defined in
the portion of the $(\tilde{C},\phi_0)$-plane, $0 \le \tilde{C}
\le \Lambda/2$ and $|\phi_0|<F_2$ (see Fig. \ref{Figure2}).

The ${\rm 1/sn}$ solution, Eq. (\ref{m1n0n0d}), is defined
for $0 \le m \le 1$, and it is complementary to the ${\rm sn}$
solution since it is also valid for $0 \le \tilde{C} \le
\Lambda/2$, but for $|\phi_0|>F_1$ (see Fig. \ref{Figure2}).

The ${\rm 1/cn}$ solution, Eq. (\ref{n0m1n0d}), is defined
for $0 \le m \le 1$ and it occupies the region $\Lambda/2 \le
\tilde{C} \le \infty$, $|\phi_0|>F_1$ (see Fig. \ref{Figure2}).

The ${\rm sndn/cn}$ solution, Eq. (\ref{p1m1p1d}), is
defined for unlimited $\phi_0$ in the range $4-8/\Lambda \le
\tilde{C} \le 0$ (see Fig. \ref{Figure2}). This solution is only
valid for $m^{\ast}<m<1$, where, for fixed $\lambda$,
$m^{\ast}(h)$ is an increasing function of $h$ and
$m^{\ast}(0)=1/2$. For $m<m^{\ast}$ the first expression in Eq.
(\ref{p1m1p1d}) does not have solutions for $\beta$. For
$m^{\ast}<m<1$, the equation has two roots, $\beta_1<\beta_2$. For
the limiting value, $m^{\ast}$, one can find the corresponding
amplitude $A^{\ast}$ from the second expression of Eq.
(\ref{p1m1p1d}), and then $\tilde{C}^{\ast}$, from the last
expression. For the case of $\Lambda=1$ presented in Fig.
\ref{Figure2}, we find $m^{\ast} \approx 0.933$ and
$\tilde{C}^{\ast} \approx -1.00$.

It should be noticed that we have not found a solution of the form
of Eq. (\ref{EllipticDiscrete}) valid in the range of $\tilde{C} <
4-8/\Lambda$ (portion marked with the question mark in Fig.
\ref{Figure2}). It is likely that static solutions in this range
cannot be expressed in terms of the Jacobi elliptic functions
because they do not survive in the continuum limit (see Sec.
\ref{ContinuumSolutions}). However, the solution can easily be
constructed from the nonlinear map in Eq. (\ref{Pulse}).

Let us now discuss further several particular examples of the above solutions.
First of all, in the limit $m \rightarrow 1$ [see Eq.
(\ref{HyperbolicDiscrete})], the ${\rm sn}$ solution, Eq.
(\ref{p1n0n0d}), reduces to the kink solution \cite{Saxena},
\begin{eqnarray}
\phi_n=\pm\tanh [\beta h(n+x_0)],
\label{KinkCooper}
\end{eqnarray}
while the ${\rm 1/sn}$ solution, Eq. (\ref{m1n0n0d}), to the
solution called hereafter the ``inverted'' kink,
\begin{eqnarray}
\phi_n=\frac{\pm 1}{\tanh [\beta h(n+x_0)]}.
\label{InvertedKink}
\end{eqnarray}
In Eqs. (\ref{KinkCooper}) and (\ref{InvertedKink}), $x_0$ is
the (arbitrary) position of the solution and $\tanh(\beta h)=\sqrt{\Lambda/2}$.

This limiting case corresponds to $\tilde{C}=0$, for which a
heteroclinic connection is possible between the fixed points
$\phi_n=-1$ and $\phi_n=1$ (see Fig. \ref{Figure2}), giving rise
to the kink or the inverted kink. In this case, Eq. (\ref{Pulse})
assumes the following simple form
\begin{eqnarray}
\phi_n=\frac{\phi_{n-1} \pm \sqrt{\Lambda/2}}{1\pm
\sqrt{\Lambda/2} \phi_{n-1}}, \label{Kink}
\end{eqnarray}
where one can choose either the upper or the lower signs.
The kink, Eq. (\ref{KinkCooper}), and the inverted kink, Eq.
(\ref{InvertedKink}), can be derived from Eq. (\ref{Kink}) taking
initial values from $|\phi_0|<1$ and $|\phi_0|>1$, respectively.

\begin{figure}
\includegraphics{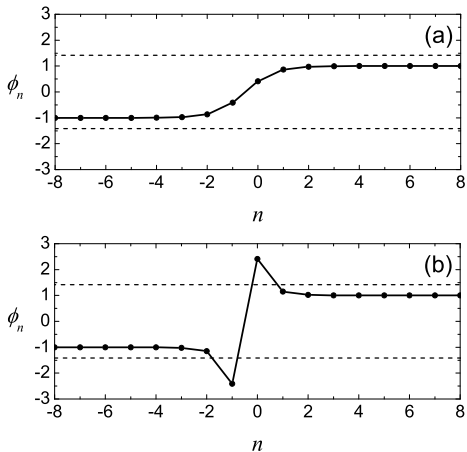}
\caption{Solutions at $\tilde{C}=0$: (a) kink and (b) inverted
kink at $\Lambda=1$. The solutions can be found from the nonlinear map
Eq. (\ref{Kink}) with the initial conditions $\phi_0=\sqrt{2}-1$
and $\phi_0=\sqrt{2}+1$ for (a) and (b), respectively.
%On the other hand, they
The kink and the inverted kink are also given by Eqs. (\ref{KinkCooper})
and (\ref{InvertedKink}) respectively, for $x_0=0.5$.
%for (a) and (b), respectively.
Horizontal dashed lines, $\phi_n= \pm \sqrt{2/\Lambda}$, show
positions of singular points of the potential $V(\phi_n)$ (see
Fig. \ref{Figure1}).} \label{Figure4}
\end{figure}

In Fig. \ref{Figure4} we show (a) the kink and (b) the inverted
kink solutions taking $\phi_0=\sqrt{2}-1$ and $\phi_0=\sqrt{2}+1$,
respectively, for $\Lambda=1$.

\begin{figure}
\includegraphics{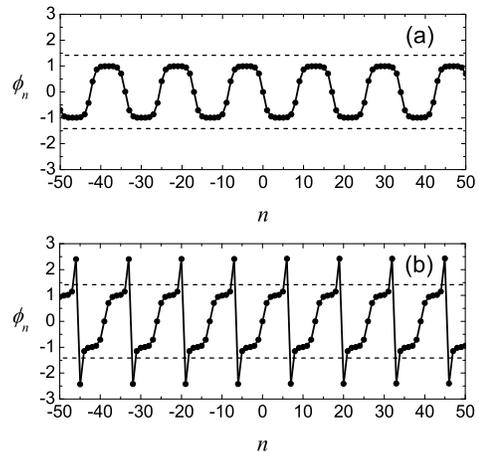}
\caption{Solutions for $\tilde{C}$ close to 0, at $\Lambda=1$,
obtained from the map of Eq. (\ref{Pulse}): (a) $\phi_0=0$,
$\tilde{C}=2\times 10^{-5}$ and (b) $\phi_0=0$,
$\tilde{C}=-6.75\times 10^{-4}$. The solution in (a) is the ${\rm
sn}$ solution, Eq. (\ref{p1n0n0d}), and the solution in (b) is the
${\rm sndn/cn}$ solution, Eq. (\ref{p1m1p1d}). Horizontal dashed
lines, $\phi_n= \pm \sqrt{2/\Lambda}$, show positions of singular
points of the potential $V(\phi_n)$ (see Fig. \ref{Figure1}).}
\label{Figure5}
\end{figure}

In Fig. \ref{Figure5}, and for $\Lambda=1$, we present two
examples of solutions for positive and negative $\tilde{C}$ close
to $0$. In particular, for $\tilde{C}=+2\times 10^{-5}$, taking
initial value $\phi_0=0$, we obtain from the map of Eq.
(\ref{Pulse}) the solution presented in panel (a). In fact, it is
the ${\rm sn}$ solution close to the hyperbolic function limit
having the form of a periodic chain of kinks and anti-kinks. On
the other hand, for $\tilde{C}=-6.75\times 10^{-4}$ and
$\phi_0=0$, we obtain from the map of Eq. (\ref{Pulse}) the
solution shown in panel (b). This is the ${\rm sndn/cn}$ solution
close to the hyperbolic function limit and has a form of a chain
of kinks and inverted anti-kinks.

\begin{figure}
\includegraphics{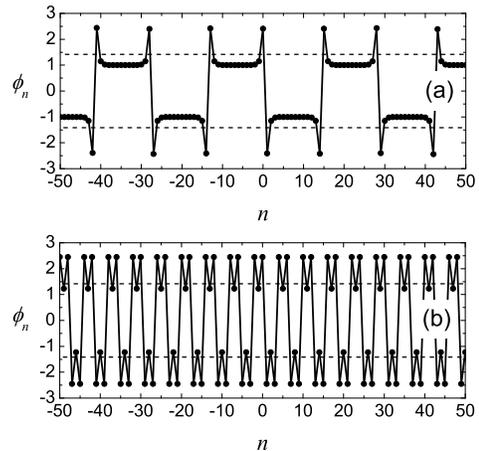}
\caption{The ${\rm 1/sn}$ solution, Eq. (\ref{m1n0n0d}), (a) close
to the hyperbolic and (b) at the trigonometric limits. We set
$\Lambda=1$. The solution in (a) consists of a chain of inverted kinks and
inverted anti-kinks. To obtain these solutions from the nonlinear
map Eq. (\ref{Pulse}) we set $\tilde{C}=1.23\times 10^{-9}$,
$\phi_0=1+\sqrt{2}$ for (a), and $\tilde{C}=0.5$, $\phi_0=2.45$ for (b).}
\label{Figure6}
\end{figure}

In Fig. \ref{Figure6}, and for $\Lambda=1$, we present the ${\rm
1/sn}$ solution, Eq. (\ref{m1n0n0d}), (a) close to the hyperbolic
and (b) at the trigonometric limits. The solution shown in panel
(a) is a chain of inverted kinks and inverted anti-kinks. These
solutions are obtained from the nonlinear map Eq. (\ref{Pulse})
setting $\tilde{C}=1.23\times 10^{-9}$, $\phi_0=1+\sqrt{2}$ for
(a), and $\tilde{C}=0.5$, $\phi_0=2.45$ for (b).

In summary, we have shown that the well-established (hyperbolic
and elliptic function) solutions of the model correspond to a very
narrow region of the two-parameter ($\phi_0,\tilde{C}$) admissible
space. A natural question is what is the typical profile outcome
stemming from the use of other pairs of $\phi_0$ and $\tilde{C}$
in Eq. (\ref{Pulse}). Generically, upon testing the different
regions of the admissible regime we have observed that arbitrary
choices may lead to seemingly erratic solutions with very large
amplitudes. A different sign choice in the right hand side of Eq.
(\ref{Pulse}) may, however, lead to a periodically locked tail
structure. A simple example of such a solution is given below.

\subsection{$\Lambda<0$ case.}

For negative $\Lambda$, we have the ${\rm cn}$ solution, Eq.
(\ref{n0p1n0d}), and the ${\rm dn}$ solution, Eq. (\ref{n0n0p1d}).

Let us first start from the latter: When $m$ increases from $0$ to $1$ in Eq. (\ref{n0n0p1d}),
the parameter $C$ decreases monotonically from $0$ to $\Lambda/2$,
i.e., this solution is defined in the ``dn'' portion of Fig.
\ref{Figure3}.

On the other hand, the ${\rm cn}$ solution of Eq. (\ref{n0p1n0d})
is only valid for $m^{\ast}<m<1$, where $m^{\ast}(h)$ is an
increasing function of $h$ and $m^{\ast}(0)=1/2$. For $m<m^{\ast}$
the second expression of Eq. (\ref{n0p1n0d}) does not have
solutions for $\beta$. For $m^{\ast}<m<1$, the equation has two
roots, $\beta_1<\beta_2$. For the limiting value, $m^{\ast}$, one
can find the corresponding amplitude $A^{\ast}$ from the second
expression of Eq. (\ref{n0p1n0d}) and then $\tilde{C}^{\ast}$ from
the last expression. When $m$ increases from $m^{\ast}$ to $1$ in
Eq. (\ref{n0p1n0d}), the parameter $\tilde{C}$ of the nonlinear
map Eq. (\ref{Pulse}) corresponding to the root $\beta_1$
($\beta_2$) increases from $\tilde{C}^{\ast}$ (decreases from
$\tilde{C}^{\ast}$) to $\Lambda/2$ (to $-\infty$). Thus, the ${\rm
cn}$ solution occupies the rest of the admissible region for the
case of $\Lambda<0$ (see Fig. \ref{Figure3}). In the case of
$\Lambda=-1$ presented in Fig. \ref{Figure3}, we find $m^{\ast}
\approx 0.873$ and $\tilde{C}^{\ast} \approx -2.96$.

\begin{figure}
\includegraphics{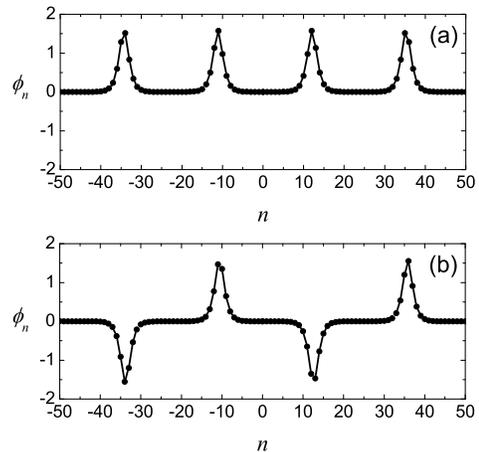}
\caption{Solutions for $\Lambda=-1$ at (a)
$\tilde{C}=\Lambda/2-10^{-8}$ and (b) $\tilde{C}
=\Lambda/2+10^{-8}$, for initial value of $\phi_{0}=10^{-4}$.
Since $\tilde{C}$ is close to $\Lambda/2=-1/2$, corresponding
to $m=1$, the dn solution in (a) and the cn solution in (b) look like
chains of separated pulses given by Eq. (\ref{PulseCooper}).}
\label{Figure7}
\end{figure}

Both ${\rm cn}$ and ${\rm dn}$ solutions, in the limit $m
\rightarrow 1$ ($\tilde{C} \rightarrow \Lambda/2$), reduce to a
homoclinic to $0$ pulse solution (see also Figs. \ref{Figure1} and
\ref{Figure3} where this possibility is illustrated). This
solution has the form:
\begin{eqnarray}
\phi_n=\pm A\,{\rm sech} [\beta h(n+x_0)],
\label{PulseCooper}
\end{eqnarray}
where
\begin{eqnarray}
\cosh(\beta h)=1-\frac{\Lambda}{2},\,\,\,\,\,\,{\rm and}\,\,\,\,\,\, %\nonumber \\
A=\sqrt{2-\frac{\Lambda}{2}}. \label{AandPHI}
\end{eqnarray}
This is illustrated by Fig. \ref{Figure7} where, taking
$\Lambda=-1$, we show the solution obtained from the map Eq.
(\ref{Pulse}) at (a) $\tilde{C}=\Lambda/2-10^{-8}$ (${\rm dn}$
solution) and (b) $\tilde{C}=\Lambda/2+10^{-8}$ (${\rm cn}$
solution), for initial value of $\phi_{0}=10^{-4}$. The figure
clearly illustrates the two limits (to the left and to the right
of $\tilde{C}=\Lambda/2$) and their correspondence to pairs of
pulse--pulse solutions and ones of pulse--anti-pulse solutions, as
one enters the two different regimes ``dn'' and ``cn'' of Fig.
\ref{Figure3}.

\subsection{Solutions with $\phi_{n-1}=\phi_{n+1}$}

Solutions of this special form can be expressed as $\phi_n =
A\cos(\pi n) +B$ with constant $A$ and $B$. For the sake of
simplicity, we set $B=0$ and substitute the ansatz into Eq.
(\ref{PhysicaD}) or Eq. (\ref{Saxena}) to find the zigzag solution
\begin{eqnarray}
\phi_n=A\cos(\pi n),\,\,\,\,\,\,\, A=\sqrt{\frac{4} {\Lambda} -1}
\,. \label{Zigzag}
\end{eqnarray}
To obtain this solution from the map Eq. (\ref{Pulse}), one has to
set $\tilde{C}=(\lambda/2)(1+A^2)^2-4A^2/h^2$, which is the
general condition for getting $\phi_{n}=-\phi_{n-1}$. Substituting
here $A$ from Eq. (\ref{Zigzag}), we get
$\tilde{C}=4(\Lambda-2)/(\Lambda h^2)$. For the case of
$\Lambda=1$ presented in Fig. \ref{Figure2}, we have
$\tilde{C}=-4$.

The zigzag solution is an exceptional one, as
$\phi_{n}=\phi_{n-2}$ for any $n$. However, as shown above,
this is only possible when Eq. (\ref{Pulse}) has multiple roots, which means
that the two-point static problem, Eq. (\ref{FirstIntDiscrete}), is factorized.
This solution does not have a counterpart in the continuum limit.

\section{Linear stability}
\label{LinearStability}

The static solutions of the discrete models Eq. (\ref{PhysicaD})
and Eq. (\ref{Saxena}) are exactly the same, but the dynamical
properties of the two models are different. As the corresponding
stability analysis takes into account the dynamical form of each
model, it will be carried out separately for each of the two.

Let us first consider the MC model of Eq. (\ref{PhysicaD}).
Introducing the ansatz $\phi_n(t)=\phi_n^0+\varepsilon_n(t)$
(where $\phi_n^0$ is an equilibrium solution and
$\varepsilon_n(t)$ is a small perturbation), we linearize Eq.
(\ref{PhysicaD}) with respect to $\varepsilon_n$ and obtain the
following equation:
\begin{eqnarray}
\ddot{\varepsilon}_n=\frac{1}{h^2}( \varepsilon_{n-1} -
2\varepsilon_{n} +\varepsilon_{n+1} ) + \lambda \varepsilon_{n}
\nonumber \\
-\frac{\lambda}{2} (\phi_n^0)^2 ( \varepsilon_{n-1}
+\varepsilon_{n+1} ) - \lambda \phi_n^0(\phi_{n-1}^0+\phi_{n+1}^0)
\varepsilon_{n}.
\label{LinPhysicaD}
\end{eqnarray}
For the small-amplitude phonons, $\varepsilon_{n}=\exp(i \kappa n
+i \omega t)$, with frequency $\omega$ and wave number $\kappa$,
around the uniform steady states $\phi_0=\pm 1$ ($\lambda>0$), Eq.
(\ref{LinPhysicaD}) is reduced to the following dispersion
relation:
%
%\begin{eqnarray}
%\omega^2=\left[\frac{\lambda}{2} (\phi_n^0)^2- \frac{1}{h^2}
%\right]
%\left( 2-4\sin^2\frac{\kappa}{2} \right) \nonumber \\
%+\frac{2}{h^2} -\lambda +\lambda \phi_n^0 (\phi_{n-1}^0 +
%\phi_{n+1}^0).
%\label{SpecPhysicaD}
%\end{eqnarray}
%
%>From Eq. (\ref{SpecPhysicaD}), the spectrum of the vacuum
%solutions for $\lambda>0$, $\phi_n^0=\pm 1$, is
%
\begin{eqnarray}
\omega^2=2\lambda + \left(\frac{4}{h^2} - 2\lambda \right)
\sin^2\left(\frac{\kappa}{2} \right),
\label{SpecVacPhysicaD}
\end{eqnarray}
while the spectrum of the vacuum solution for $\lambda<0$, $\phi_n^0=0$, is
\begin{eqnarray}
\omega^2= \frac{4}{h^2} \sin^2\left(\frac{\kappa}{2}
\right)-\lambda. \label{SpecVacPhysicaDN}
\end{eqnarray}

For an arbitrary stationary solution $\phi_n^0$, stability
is inferred in the MC model if
the eigenvalue problem obtained from Eq. (\ref{LinPhysicaD}) by
replacing $\ddot{\varepsilon}_n$ with $-\omega^2\varepsilon_n$ has
only non-negative solutions $\omega^2$. Recalling that the MC
model is a non-Hamiltonian one, the resulting eigenvalue problem is a
non-self-adjoint one involving a non-symmetric
matrix.

Similarly, we obtain analogous expressions for the EC model of
Eq. (\ref{Saxena}).
The linearized equation reads
\begin{eqnarray}
\ddot{\varepsilon}_n &=&\frac{1}{h^2}( \varepsilon_{n-1}
- 2\varepsilon_{n}+\varepsilon_{n+1} ) \nonumber \\
&+& 2\lambda\frac {2+\left(\Lambda -6 \right)(\phi_n^0)^2 + \Lambda
(\phi_n^0)^4} {\left[ 2- \Lambda (\phi_n^0)^2 \right]^2}
\varepsilon_{n},
\label{LinSaxena}
\end{eqnarray}
and the corresponding dispersion relation for the linear phonon modes
around $\phi_n^0= \pm 1$
has the form:
%
%\begin{eqnarray}
%\omega^2=\frac{4}{h^2}\sin^2 \left(\frac{\kappa}{2} \right)
%\nonumber \\
%-2\lambda\frac{2+ \left( \Lambda -6\right)(\phi_n^0)^2 +\Lambda
%(\phi_n^0)^4} {\left[ 2-\Lambda (\phi_n^0)^2 \right]^2}.
%\label{SpecSaxena}
%\end{eqnarray}
%
%Hence, the spectrum of vacuum solutions $\phi_n=\pm 1$ (for $\lambda>0$) is
%
\begin{eqnarray}
\omega^2=\frac{4\lambda}{2-\Lambda } + \frac{4}{h^2} \sin^2
\left(\frac{\kappa}{2} \right),
\label{SpecVacSaxena}
\end{eqnarray}
while that of vacuum solution $\phi_n=0$ (for $\lambda<0$) is
\begin{eqnarray}
\omega^2=\frac{4}{h^2}\sin^2 \left(\frac{\kappa}{2} \right)-
\lambda \,.
\label{SpecVacSaxenaN}
\end{eqnarray}

The stationary solution $\phi_n^0$ is stable in the EC model of
Eq. (\ref{Saxena}) if the self-adjoint eigenvalue problem obtained
from Eq. (\ref{LinSaxena}) by replacing $\ddot{\varepsilon}_n$
with $-\omega^2\varepsilon_n$ has only non-negative solutions
$\omega^2$.

We note that all stable and unstable static solutions of the MC
and the EC models, except for the zigzag solution Eq.
(\ref{Zigzag}), possess a zero-frequency mode. This is a
consequence of the effective translational invariance of the
discrete PNb-free models; this is related also to the freedom of
selecting the free parameter $x_0$ in the corresponding solution
expressions.

Our aim here is not to study the whole bunch of solutions in the
whole range of parameters, but rather to demonstrate the existence
of stable solutions and also provide some examples where
solutions, being stable in one model, may be unstable in another.
The results of stability analysis for some characteristic
solutions are summarized in Table \ref{Table}.

First, we consider the kink and inverted kink solutions given in
Eq. (\ref{KinkCooper}) and  Eq. (\ref{InvertedKink}) respectively:
In a numerical experiment, these solutions were placed in the
middle of a chain of $N=200$ particles with clamped boundary
conditions. We have found that for the chosen set of parameters,
the kink is stable in both MC and EC models, while {\it the
inverted kink is stable in the EC and unstable in the MC model}.

\begin{figure}
\includegraphics{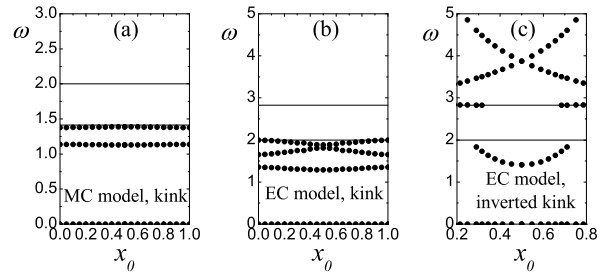}
\caption{Spectrum of (a) kink in MC model, (b) kink in EC model,
and (c) inverted kink in EC model at $\tilde{C}=0$, $\Lambda=1$
($h=1$). The inverted kink in MC model is unstable for the chosen
parameters. Horizontal lines show the borders of the phonon
bands.} \label{Figure8}
\end{figure}

In Fig. \ref{Figure8} we show the spectra of the kink and the inverted
kink for their different positions with respect to the lattice, $x_0$.
In the figure, the horizontal lines show the borders of the
phonon band of the vacuum $\phi_n=\pm 1$ and the dots show the
frequencies of the kink's internal modes lying outside of the
band. In all cases, the spectra are shown as functions of $x_0$,
even though, admittedly, the kink in the MC model presented in (a)
demonstrates very weak sensitivity of its spectrum to variations of $x_0$.

Let us now consider the periodic solutions depicted in Fig.
\ref{Figure5}(b) (${\rm sndn/cn}$) and in Fig. \ref{Figure6}(a)
(${\rm 1/sn}$). For these classes of solutions we used periodic boundary
conditions and the length of the lattice was commensurate with the
period of the solutions containing a number of periods. Close to the
hyperbolic function limit, these solutions are stable in the EC
model but unstable in the MC model, following the stability of the
building blocks (inverted kinks) associated with these structures. On the
other hand, the ${\rm 1/sn}$ solution becomes unstable in both
models at the trigonometric limit depicted in Fig.
\ref{Figure6}(b). Nonetheless, we find it remarkable that some of
these ``apparently non-smooth'' solutions (or maybe more
correctly, solutions apparently containing a non-smooth continuum
limit) may be potentially stable in the discrete setting. It would
be wortwhile to analyze the stability of such structures in more
detail and as a function of their elliptic modulus $m$.

Next, we consider the zigzag solution, Eq. (\ref{Zigzag}), which
exists for $0<\Lambda<4$. It can be shown that this solution can
be stable in the EC model: Indeed, substituting Eq. (\ref{Zigzag})
into the relevant eigenvalue problem, we obtain the dispersion relation for
the linear phonon modes, $\omega^2= (4/h^2) \sin^2(\kappa/2) +8 /
[h^2(2-\Lambda)]$. The solution is stable when $\omega^2$ are
non-negative, i.e., when $\Lambda<2$. Combining this with the
existence condition we find that the zigzag solution exists and it
is stable in the EC model for $0<\Lambda<2$. Note that this
exceptional solution does not possess the zero frequency mode, as
it can be seen from the above dispersion relation. The amplitude
of the solution, $\sqrt{4/\Lambda-1}$, is always greater than 1.
The zigzag solution is always unstable in the MC model, a result
that can similarly be demonstrated.

The ${\rm sn}$, ${\rm cn}$, and ${\rm dn}$ solutions close to the
hyperbolic limit were found to be generically unstable in both the
MC and EC models.
The instability of the pulse solution of Eq. (\ref{PulseCooper})
can be analytically justified in the present setting. It can be
easily inferred from the invariance of the solution with respect
to $x_0$ that the eigenvector leading to the zero eigenvalue of
Fig. \ref{Figure8} is proportional to $\partial \phi/\partial x_0$
(where $\phi$ is the pulse solution profile). This is an
anti-symmetric eigenvector, given the pulse's symmetric (around
its center) nature. But then, from Sturm-Liouville theory for
discrete operators \cite{levy}, there should be an eigenvalue such
that $\omega^2<0$ with a symmetric (i.e., even) eigenvector,
resulting in the instability of the relevant pulse solution. Our
numerical simulations fully support this conclusion. The pulse
solution was constructed by iterating Eq. (\ref{Pulse}) for
$\Lambda<0$, $\tilde{C}=\Lambda/2$, and $0<\phi_{0}<\sqrt{2-\Lambda/2}$,
or directly from Eq. (\ref{PulseCooper}).
We found that for $\lambda=-1$ the pulse solution is unstable
in both models and over a wide range of the discreteness parameter $h$,
including the case of rather small $h\approx 0.1$. We have also confirmed
that the instability mode is similar to the pulse profile (i.e., of even parity).

Both the cn and dn solutions (see, e.g., Fig. \ref{Figure7})
are also found to be unstable for the different
parameter values that we have used in both the MC and EC models.
This is rather natural to expect given that their ``building
blocks'', namely the pulse-like solutions are each unstable (hence
their concatenation will only add to the unstable eigendirections).

Finally, as far as the sn solutions are concerned, two basic
instability modes were revealed. One resembles the instability of
the pulse solution, when the whole structure, being waved around
the top of the background potential, tends to slide down to one of
the potential wells. The other instability modes can be understood
if one regards the sn solution as a set of alternating kinks and
anti-kinks [see Fig. \ref{Figure5}(a)] interacting with each other
by means of overlapping tails. Such a system is unstable because
the kinks tend to annihilate with the anti-kinks.

It is interesting to note that these stability results suggest a
disparity with the earlier numerical findings of \cite{Saxena}
where the pulse-like solution was generically reported to be
stable and analogous statements were made for the cn- and dn-type
solutions.

At this point, it is worth discussing a connection with NLS-type
models mentioned in the introduction. In the case of such models,
the aforementioned unstable eigendirection (with an eigenvector
similar in profile to the pulse itself) is ``prohibited'' by the
additional conservation law of the $l^2$ norm, resulting in a
separate neutral eigendirection with zero eigenfrequency. As a
consequence, it is an interesting twist that the instability
reported for such pulse-like solutions in scalar $\phi^4$ models
would be absent in their discrete NLS counterparts.

\begin{table}
\caption{Results of stability analysis \label{Table}}
\begin{ruledtabular}
\begin{tabular}{|l|c|c|}
  Solution    & MC model,  & EC model,   \\
      &  Eq. (\ref{PhysicaD})   & Eq. (\ref{Saxena})  \\ \hline
  kink, Eq. (\ref{KinkCooper})    & stable & stable \\ \hline
  inverted kink,    & unstable & stable \\
  Eq. (\ref{InvertedKink})  &   &  \\ \hline
  pulse, Eq. (\ref{PulseCooper})   & unstable & unstable \\ \hline
  {\rm sn \,\,cn,\,\,dn} & unstable & unstable \\
  close to hyper- &   &   \\
  bolic limit &   &   \\ \hline
  {\rm sndn/cn,\,\,1/sn} & unstable & stable \\
  close to hyper- &   &   \\
  bolic limit &   &   \\
\end{tabular}
\end{ruledtabular}
\end{table}

\section{Slow kink dynamics}
\label{Dynamics}

In \cite{JPhysA}, we have compared the spectra and the long-term
dynamics of the kinks in the two PNb-free models, Eq.
(\ref{PhysicaD}) and Eq. (\ref{Saxena}), with that of the
classical discrete model, Eq. (\ref{KleinGordonDiscrete}) (see
Fig. 1 of that paper). The case of very small kink velocities was
not studied there. Here we would like to focus on this case to
demonstrate the qualitatively different behavior for slowly moving
kinks in the PNb-free and classical models.

To boost the kink in the PNb-free models Eq. (\ref{PhysicaD}) and
Eq. (\ref{Saxena}) we used the following dynamical solution
corresponding to the multiple eigenvalue $\omega^2=0$,
$\phi_n(t)=\phi_n^0 + ct\varepsilon_n$, where $\phi_n^0$ is a static
kink solution, $\varepsilon_n$ is the normalized translational
eigenmode and $c$ is the amplitude playing the role of kink
velocity. This construction yields a more accurate approximate
solution for very small $c$ when linearized equations Eq.
(\ref{LinPhysicaD}) and Eq. (\ref{LinSaxena}) are accurate. Increasing
$c$ leads to a decrease in the accuracy of the dynamical solution used for
boosting. On the other hand, we note in passing that
there are techniques developed recently for Klein-Gordon
\cite{aigner,oxt1} and even nonlinear Schr{\"o}dinger lattices
\cite{preprint} that also allow the construction of finite speed,
numerically exact travelling solutions in these classes of models.

\begin{figure}
\includegraphics{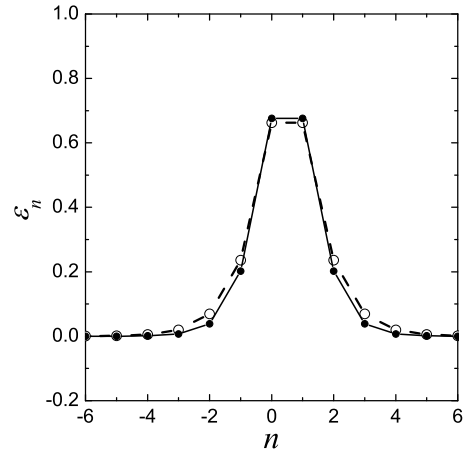}
\caption{Lowest frequency normalized eigenmodes for inter-site
kinks. Both PNb-free models, Eq. (\ref{PhysicaD}) and Eq.
(\ref{Saxena}), have the same translational eigenmode with
$\omega=0$, shown by dots connected with solid lines (see  also
the relevant discussion in the text). For the classical model, Eq.
(\ref{KleinGordonDiscrete}), the lowest-frequency mode has $\omega
\approx 0.252$ (shown by open circles and dashed lines). Results
for $h=1$, $\lambda=1$ ($\Lambda=1$).} \label{Figure9}
\end{figure}

In Fig. \ref{Figure9} we present the lowest frequency normalized
eigenmodes for the inter-site kinks in the three models, for the
case of $h=1$, $\lambda=1$ ($\Lambda=1$). Both PNb-free models,
Eq. (\ref{PhysicaD}) and Eq. (\ref{Saxena}), have the same
translational eigenmodes with $\omega=0$, shown by dots connected
with solid lines. One can show by straightforward algebra that the
PNb-free model modified by a non-vanishing multiplier
$e(h,\phi_{n-1},\phi_n,\phi_{n+1})$ as described in Sec.
\ref{sec:Generalizations} has the same translational eigenmode as
the original model. Assuming that the two models are respectively:
\begin{eqnarray}
\ddot{\phi}_n &=& f(\phi_{n-1},\phi_n,\phi_{n+1})
\label{extra1}
\\
\ddot{\phi}_n &=& e(\phi_{n-1},\phi_n,\phi_{n+1}) \times
f(\phi_{n-1},\phi_n,\phi_{n+1}),
\label{extra2}
\end{eqnarray}
where the function $e(x,y,z) \neq 0$, then the steady
state solutions $\phi_n^0$ will satisfy $f(\phi_{n-1}^0,\phi_n^0,\phi_{n+1}^0)
=0$, for all $n$, while the corresponding linearization equations
(using $\phi_n=\phi_n^0 + \epsilon_n$)
are respectively of the form
\begin{eqnarray}
\ddot{\epsilon}_n &=& \left. \frac{\partial f}{\partial
\phi_{n-1}}\right|_0 \epsilon_{n-1}
                    + \left. \frac{\partial f}{\partial \phi_{n}}\right|_0 \epsilon_{n}
                    +  \left. \frac{\partial f}{\partial \phi_{n+1}}\right|_0
                    \epsilon_{n+1},
\label{extra3}
\\
\ddot{\epsilon}_n &=& e \left[\left. \frac{\partial f}{\partial
\phi_{n-1}}\right|_0 \epsilon_{n-1}
                    + \left. \frac{\partial f}{\partial \phi_{n}}\right|_0 \epsilon_{n}
                    +  \left. \frac{\partial f}{\partial \phi_{n+1}}\right|_0
                    \epsilon_{n+1}
\right]. \label{extra4}
\end{eqnarray}
Hence, when solving the corresponding eigenvalue problem (again
substituting $\ddot{\epsilon}_n$ by $-\omega^2 \epsilon_n$), for
$\omega=0$, the eigenvalue problems become identical, both satisfying
\begin{eqnarray}
0=\left. \frac{\partial f}{\partial \phi_{n-1}}\right|_0 \epsilon_{n-1}
                    + \left. \frac{\partial f}{\partial \phi_{n}}\right|_0 \epsilon_{n}
                    +  \left. \frac{\partial f}{\partial \phi_{n+1}}\right|_0
\epsilon_{n+1}, \label{extra5}
\end{eqnarray}
hence the coincidence of the corresponding eigenvectors. We are
extremely grateful to an anonymous referee for this remark and its
proof. We do note, in passing, also that while this justifies the
coincidence of the zero eigenfrequency modes of the PNb-free
models in Fig. \ref{Figure9}, it is also in tune with the results
of Fig. \ref{Figure8} for nonzero eigenfrequencies $\omega$. The
latter are not identical between the different models, as
indicated by the left and middle panels of that figure. This is
due to the differences between the corresponding eigenvalue
problems of Eqs. (\ref{extra3}) and (\ref{extra4}), when $\omega
\neq 0$.

For the classical model, Eq. (\ref{KleinGordonDiscrete}), the
lowest-frequency mode has $\omega \approx 0.252$, and it is shown
in Fig. \ref{Figure9} by open circles and dashed lines. Actually,
this mode is not a translational mode (since, strictly speaking,
there is no translational invariance) but we use it to boost the
kink. One can say that this will {\it become} the translational
mode for this model in the continuum limit.

We define the kink center of mass as
\begin{eqnarray}
S=\frac{\sum_n n\sqrt{1-\phi_n^2}}{\sum_n \sqrt{1-\phi_n^2}}.
\label{KinkPosition}
\end{eqnarray}

\begin{figure}
\includegraphics{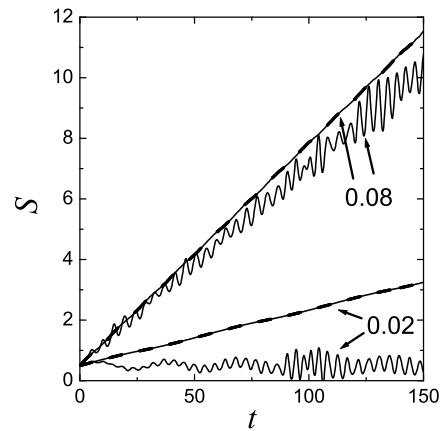}
\caption{The kink coordinate, $S$, as a function of time for kinks
propagating in the PNb-free MC (dashed line) and EC (solid line)
models, as well as in the classical discrete $\phi^4$ model
(oscillatory line). The kinks were boosted with two different
amplitudes of the normalized lowest-frequency eigenmodes, $c=0.02$
and $c=0.08$. The faster classical kink is able to propagate,
while the slower one is not, since it cannot overcome the PN
barrier. Kinks in PNb-free models are not trapped by the lattice
and can therefore propagate.} \label{Figure10}
\end{figure}

The evolution of the kink coordinates is shown in Fig.
\ref{Figure10}. Kinks were boosted with two different amplitudes
of the normalized lowest-frequency eigenmodes, $c=0.02$ and
$c=0.08$. Results for the PNb-free models practically coincide, as
depicted by the dashed and solid lines for the MC and EC models,
respectively. It is readily seen that kinks in the PNb-free
lattices propagate with roughly constant velocities.

The oscillatory trajectories in Fig. \ref{Figure10} correspond to
the kink in the classical discrete $\phi^4$ model. The faster
kink, boosted with $c=0.08$, propagates along the lattice but its
velocity gradually decreases. The kinetic energy of the
translational motion is partly lost to the excitation of the kink
internal mode with $\omega \approx 1.26$, lying below the phonon
frequency band. Higher harmonics of this mode interact with the
phonon spectrum producing radiation, which also slows the kink
down. An even more dramatic difference is observed for the slower
kinks, boosted with $c=0.02$. Here, the classical kink cannot
overcome the PN barrier and can not propagate, oscillating near
the stable inter-site configuration. On the other hand, the kinks
in the PNb-free models are not trapped by the lattice and
propagate due to the absence of the PN barrier. Alternatively, one
can say that such waveforms can be accelerated by arbitrarily
small external fields.

\section{Solutions for continuum $\phi^4$ field}
\label{ContinuumSolutions}

In the continuum limit, $h \rightarrow 0$, the borders of the
admissible region, Eq. (\ref{Borders}), become
\begin{eqnarray}
(\phi_{0}^2)_{1,2}=1 \pm\sqrt{ \frac{2C }{\lambda}}.
\label{BordersCont}
\end{eqnarray}

\begin{figure}
\includegraphics{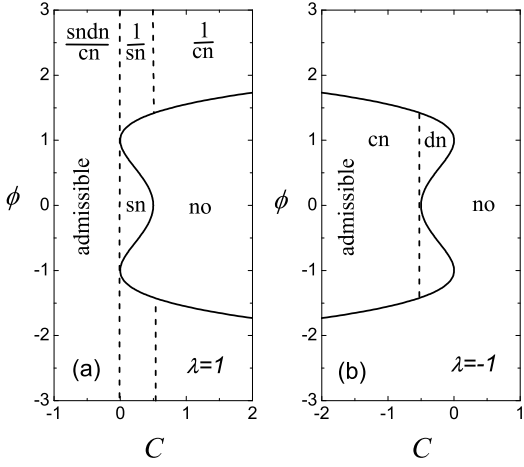}
\caption{Admissible regions for the $\phi^4$ field for (a)
$\lambda=+1$ and (b) $\lambda=-1$ obtained as the continuum limits
($h \rightarrow 0$) of those presented in Fig. \ref{Figure2} and
Fig. \ref{Figure3}, respectively. In each panel there is one
inadmissible region marked with ``no''.} \label{Figure11}
\end{figure}

In Fig. \ref{Figure11} we plot the admissible regions for (a)
$\lambda=1$ and (b) $\lambda=-1$. The topology of the admissible
regions for the $\phi^4$ field is simpler than the one pertaining
to the discrete models. In the continuum limit there exists a sole
inadmissible region for both cases $\lambda = \pm 1$, while three
inadmissible regions exist for the discrete models in the case
$\Lambda>0$. One more simplification is that the domains of the
${\rm sndn/cn}$ and ${\rm cn}$ solutions do not split into two
parts since the smaller root $\beta_1$ disappears in the continuum
limit. Particularly we note that the region marked with the
question mark in Fig. \ref{Figure2}, $\tilde{C}<4-8/\Lambda$,
disappears in the continuum limit. That might be the reason why we
failed to find a Jacobi elliptic function expression for the
discrete static solutions in this case.

Static solutions Eqs. (\ref{p1n0n0d}-\ref{p1m1p1d}) obtained for
the discrete models have their continuum counterparts as the
traveling solutions to the $\phi^4$ field, Eq.
(\ref{KleinGordon}). The general form of the solutions is
\begin{eqnarray}
\phi(x,t)&=&\pm A {\rm sn}^q(z,m) {\rm cn}^r(z,m) {\rm dn}^s(z,m) ,
\nonumber \\
z&=&\beta \left(x+x_0- ct\right), \label{EllipticContinuum}
\end{eqnarray}
where $0 \le m \le 1$ is the modulus of the Jacobi elliptic
functions, $A$ and $\beta$ are the parameters of the solution,
$x_0$ is the arbitrary initial position and $0 \le c < 1$ is the
velocity of the solution. The integers $q,r,s$ once again specify
the particular form of the solution.

Continuum analogues of Eqs. (\ref{p1n0n0d})-(\ref{p1m1p1d}) have the
following form and are characterized by the following parameters:

The ${\rm sn}$ solution, $(q,r,s)=(1,0,0)$,
\begin{eqnarray}
\beta = \sqrt{\frac{\lambda}{(1+m)(1-c^2)}}, \,\,\,\,\,\,\,\,
A=\sqrt{\frac{2m}{1+m}}, \nonumber \\
C=\frac{\lambda}{2} \left(1 - \frac{A^4}{m} \right),
\,\,\,\,\,\,\,\,\,\,\, 0<C< \frac{\lambda}{2}.
\label{p1n0n0}
\end{eqnarray}

The ${\rm cn}$ solution, $(q,r,s)=(0,1,0)$,
\begin{eqnarray}
\beta = \sqrt{\frac{-\lambda}{(2m-1)(1-c^2)}}, \,\,\,\,\,\,\,\,
A=\sqrt{\frac{2m}{2m-1}}, \nonumber \\
C=\frac{\lambda}{2} (1-A^2)^2, \,\,\,\,\,\,\,\,\,\,\, -\infty<C<
\frac{\lambda}{2}.
\label{n0p1n0}
\end{eqnarray}

The ${\rm dn}$ solution, $(q,r,s)=(0,0,1)$,
\begin{eqnarray}
\beta = \sqrt{\frac{-\lambda}{(2-m)(1-c^2)}}, \,\,\,\,\,\,\,\,
A=\sqrt{\frac{2}{2-m}}, \nonumber \\
C=\frac{\lambda}{2} (1-A^2)^2, \,\,\,\,\,\,\,\,\,\,\,
\frac{\lambda}{2} <C<0.
\label{n0n0p1}
\end{eqnarray}

The ${\rm 1/sn}$ solution, $(q,r,s)=(-1,0,0)$,
\begin{eqnarray}
\beta = \sqrt{\frac{\lambda}{(1+m)(1-c^2)}}, \,\,\,\,\,\,\,\,
A=\sqrt{\frac{2}{1+m}}, \nonumber \\
C=\frac{\lambda}{2}(1 - m A^4), \,\,\,\,\,\,\,\,\,\,\,
0<C<\frac{\lambda}{2}.
\label{m1n0n0}
\end{eqnarray}

The ${\rm 1/cn}$ solution, $(q,r,s)=(0,-1,0)$,
\begin{eqnarray}
\beta = \sqrt{\frac{\lambda}{(1-2m)(1-c^2)}}, \,\,\,\,\,\,\,\,
A=\sqrt{\frac{2(1-m)}{1-2m}}, \nonumber \\
C=\frac{\lambda}{2}\left(1+ \frac{mA^4} {1-m} \right),
\,\,\,\,\,\,\,\,\,\,\, \frac{\lambda}{2}<C< \infty.
\label{n0m1n0}
\end{eqnarray}

The ${\rm sndn/cn}$ solution, $(q,r,s)=(1,-1,1)$,
\begin{eqnarray}
\beta = \sqrt{\frac{\lambda}{2(2m-1)(1-c^2)}}, \,\,\,\,\,\,\,\,
A=\frac{1}{\sqrt{2m-1}}, \nonumber \\
C=\frac{\lambda}{2}\left(1- A^4 \right), \,\,\,\,\,\,\,\,\,\,\,
-\infty <C< 0.
\label{p1m1p1}
\end{eqnarray}

The above six solutions can be rewritten in a great variety of
forms using the properties of the Jacobi elliptic functions
\cite{SpecialFunctions}. However, we believe that they may
constitute the full list of physically different solutions to the
continuum $\phi^4$ equation since they fill the whole
two-parameter space, $(C,\phi)$, obtained as the continuum limit
of corresponding space of the discrete models.

All solutions are conveniently parameterized by a single parameter
$-\infty < C < \infty$ for $\lambda>0$, and $-\infty < C \le 0$
for $\lambda<0$, as it is presented in Fig. \ref{Figure11}.

The solutions in Eqs. (\ref{p1n0n0})-(\ref{n0n0p1}) are bounded while the
other ones are unbounded. The solutions in Eq. (\ref{n0p1n0}) and Eq.
(\ref{n0n0p1}) are defined for $\lambda <0$ while the others for
$\lambda>0$. The solutions in Eq. (\ref{n0p1n0}) and Eq.
(\ref{p1m1p1}) are valid for $1/2 < m \le 1$, the solution in Eq.
(\ref{n0m1n0}) is valid for $0 \le m < 1/2$, while the other
solutions for $0 \le m \le 1$.

\section{Conclusions}
\label{Conclusions}

In the present paper we have shown that the reduction of the
static problem of a discrete Klein-Gordon (and by extension of the
standing wave problem of a discrete NLS) equation to a two-point
problem is a powerful tool for obtaining all possible static
solutions of the corresponding model. We have applied this general
idea to a momentum conserving \cite{PhysicaD} and an energy
conserving \cite{Saxena} discretization of the $\phi^4$ model
analyzing the full two-parameter plane of solutions and giving a
natural parametrization for it. In particular, we have examined
the admissible regions of the field value at a given point and of
the constant entering the two-point function pertinent to the
model. We have specifically illustrated how to use different
choices of these two parameters to obtain not only the well-known
hyperbolic function solutions and the established elliptic
function solutions, but also novel classes of solutions including
the inverted (non-monotonic) kink solutions presented herein and
multi-kink generalizations thereof. We performed such computations
both for the attractive (focusing) and for the repulsive
(defocusing) types of nonlinearity.

The presented methodology has the significant advantage over
earlier work that by introducing the integration constant in the
discretization of the first integral of the static part of the
PDE, it allows one to construct the {\it full} family of the
static solutions of the corresponding model. Earlier work had
implicitly set this additional free parameter to $0$; this choice
was sufficient for obtaining the important hyperbolic function
solutions, but the present formalism systematically illustrates
how to generalize the latter.

We have also derived the continuum analogues of the discrete solutions
and found that they fill the whole space of parameters obtained in
the limit of $h\rightarrow 0$ from the parameter space of the
discrete models.

In addition, we have examined some of the key stability features
of the solutions obtained in the various models, illustrating the
different stability properties of the models considered herein
(even if their static solutions are identical). We have obtained
interesting stability properties, including, e.g., the
counter-intuitive stability of the inverted kink and of some of
its periodic generalizations for the EC model.

The present study generates a variety of interesting questions.
For instance, it would be relevant to examine whether the
solutions presented herein (including the inverted ones) have
counterparts in the ``standard'' discrete $\phi^4$ model and to
analyze their respective stability. It would also be relevant to
examine how the stability (and existence --since some of them may
not survive the continuum limit--) of such solutions depend on the
lattice spacing $h$. Finally, these and related (e.g. stability)
questions in the context of discrete NLS lattices (see
\cite{DNLSE1,DNLSE2,DEPelinovsky}) would also be equally or even
more (given the multitude of relevant applications of the latter
model) interesting to answer. Such studies are currently in
progress and will be reported elsewhere.

\section*{Acknowledgements}
The authors would like to thank D.~E.~Pelinovsky for a number of
stimulating and insightful discussions.
We are also grateful to an anonymous referee for
offering a proof that the MC and EC models have the same
translational eigenmodes. PGK gratefully acknowledges support
from NSF-DMS-0204585, NSF-DMS-0505663 and NSF-CAREER.

\end{document}